\documentclass[a4paper,11pt]{article}

\usepackage{ifpdf}

\usepackage{epsfig,multicol,amsmath}
\usepackage[T1]{fontenc}

\usepackage{jheppub}
\usepackage{epstopdf}

\usepackage{bm}  
\usepackage{amsmath}     
\usepackage{epsfig,epsf}
\usepackage{graphicx}
\usepackage{rotating}
\usepackage{verbatim}

\def\beq{\begin{equation}}
\def\eeq{\end{equation}}
\def\bea{\begin{eqnarray}}
\def\eea{\end{eqnarray}}

\def\eq#1{{Eq.~(\ref{#1})}}
\def\fig#1{{Fig.~\ref{#1}}}
\newcommand{\bas}{\bar{\alpha}_S}
\newcommand{\as}{\alpha_S}

\newcommand{\Lb}{\left(}
\newcommand{\Rb}{\right)}
\parskip=\itemsep               

\newcommand{\nn}{\nonumber}

\newcommand{\h}{\frac{1}{2}}
\newcommand{\qu}{\frac{1}{4}}

\newcommand{\rv}{\vec{r}}

\newcommand{\Rv}{\vec{R}}
\newcommand{\kv}{\vec{k}_T}
\newcommand{\Qv}{\vec{Q}_T}

\newcommand{\ph}{\varphi}

\title{ Azimuthal angle correlations at large rapidities: revisiting 
 density variation mechanism}
\author[a]{E. ~Gotsman,}
\author[a,b]{~~and~~ E.~ Levin}

\affiliation[a]{Department of Particle Physics, School of Physics and Astronomy,
Raymond and Beverly Sackler
 Faculty of Exact Science, Tel Aviv University, Tel Aviv, 69978, Israel}
\affiliation[b]{Departemento de F\'isica, Universidad T\'ecnica Federico
 Santa Mar\'ia, and Centro Cient\'ifico-\\
Tecnol\'ogico de Valpara\'iso, Avda. Espana 1680, Casilla 110-V,
 Valpara\'iso, Chile}
\emailAdd{gotsman@post.tau.ac.il}
\emailAdd{leving@post.tau.ac.il, eugeny.levin@usm.cl}

\abstract{ In the paper we discuss the angular correlation  present in 
hadron-hadron collisions
 at large
 rapidity difference ($\bas\,y_{12}\,\gg\,1$). We find that  in the 
CGC/saturation approach
 the 
largest 
 contribution stems from  the density variation mechanism. 
 Our  principal results  are that the odd Fourier 
harmonics
 ($v_{2n+1}$), decrease substantially as function of $y_{12}$, while the
 even harmonics ($v_{2n}$ ),     increase  considerably  with a growth of
 $y_{12}$.
}


\preprint{TAUP - 000/17\\
\today}

\begin{document}
\maketitle
\flushbottom

\section{Introduction}
In this paper we address the problem of the azimuthal angle correlations
 of two hadrons with transverse momenta $\vec{p}_{T1}$ and
 $\vec{p}_{T2}$ and rapidities $y_1$ and $y_2$, at large values
 of $y_{12} \equiv | y_1 - y_1| \gg 1/\bas$. Our main theoretical
 assumption is that these correlations stem from  interactions
 in the initial state.  We are aware that, unlike rapidity correlations
 which at large rapidities are originated from the initial state
 interactions due to causality reasons\cite{CAUSALITY}, a substantial
 part of these correlations could  be due to the interactions in the 
final
 states\cite{FINSTATE}.
On the other hand, it has been demonstrated that at  small rapidity
 difference  $\bas\,y_{12} \,<\,1$ the interactions in the initial
 state \cite{RAJUREV,KOLU1,KOWE,KOLUCOR,GLMT,GLMBEH,GOLE,KOLULAST,GLP}
   gives the value of the correlations, which  describe the major part
 of the experimentally observed correlations\cite{CMSPP,STARAA,PHOBOSAA,
STARAA1,CMSPA,CMSAA, ALICEAA, ALICEPA,ATLASPP,ATLASPA,ATLASAA}.

\begin{figure}[ht]
    \centering
  \leavevmode
      \includegraphics[width=16cm]{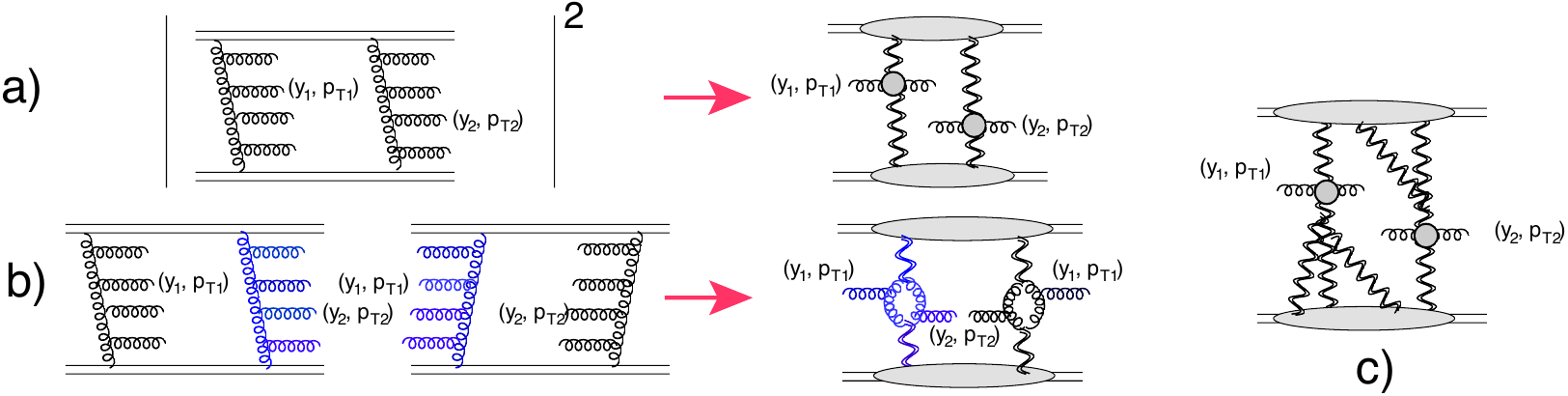}  
      \caption{ Mueller diagrams\cite{MUDIA} for two partonic showers
  production. \fig{becor}-a describes the square of the  production
 amplitudes, while \fig{becor}-b corresponds to the interference diagram
 which leads to the Bose-Einstein correlations. The wavy lines show the
 BFKL Pomeron\cite{BFKL,LI}, while the helical lines denote  gluons.
 \fig{becor}-c shows the example of  a more complicated structure of the
 partonic cascades, than the exchange of the BFKL Pomeron. The colour of
 the lines indicates  the parton shower.}
\label{becor}
   \end{figure}

 In this paper we concentrate our efforts,  on  calculating the long range rapidity  part of angular
 correlations  with large value of 
 the rapidity 
 difference $y_{12}$.  All previous calculations, 
assumed
 that $\bas y_{12}\, < \,1$\cite{RAJUREV,KOLU1,KOWE,KOLUCOR,GLMT,GLMBEH,
GOLE,KOLULAST,GLP}. It turns out, that in this kinematic region,   the main
 source of the azimuthal angle correlations, is the Bose-Einstein 
correlations
 of  identical gluons, which corresponds to the interference diagram in 
the
 production of  two partonic showers.  Intuitively, we expect that
 the correlations in the process, where two different gluons are 
produced from two
 different partonic showers,  should not depend on the difference of 
rapidities ($y_{12}$), as well as on the values of  $y_1$ and $y_2$. 
 Using the AGK cutting rules \cite{AGK}\footnote{
   In the framework 
of perturbative 
QCD  for the
 inclusive cross
 sections, the AGK cutting rules were discussed and proven in 
Refs.\cite{KOLEB,AGK1,AGK2,AGK3,AGK4,AGK5,AGK6,AGK7}. However, in
 Ref.\cite{AGK8} it
 was shown that the AGK cutting rules are violated for  double inclusive
 production. This violation is intimately related to the enhanced diagrams
 \cite{AGK7,AGK8,AGK4} and to the production of gluon from triple Pomeron vertex.  It reflects the fact that different cuts of
 the triple BFKL 
Pomeron
 vertex  with produced gluon,  lead to different contributions. Recall, that we do not
 consider such diagrams.}
 one can prove that the two gluon correlations   can be calculated
 using the Mueller diagrams\cite{MUDIA} of \fig{becor}.

 The diagrams of \fig{becor} lead to  correlations which do
 not depend on $y_1$ and $y_2$, but 
only for $\bas\,y_{12}\, \ll \,1$.  For large $y_{12}$ the contributions
 of \fig{becor}  decrease.  The main goal of this paper to find the
 contributions which survive at large $y_{12}$ ($\bas\, y_{12}\, \gg \,1$).

\begin{figure}[ht]
    \centering
  \leavevmode
      \includegraphics[width=18cm]{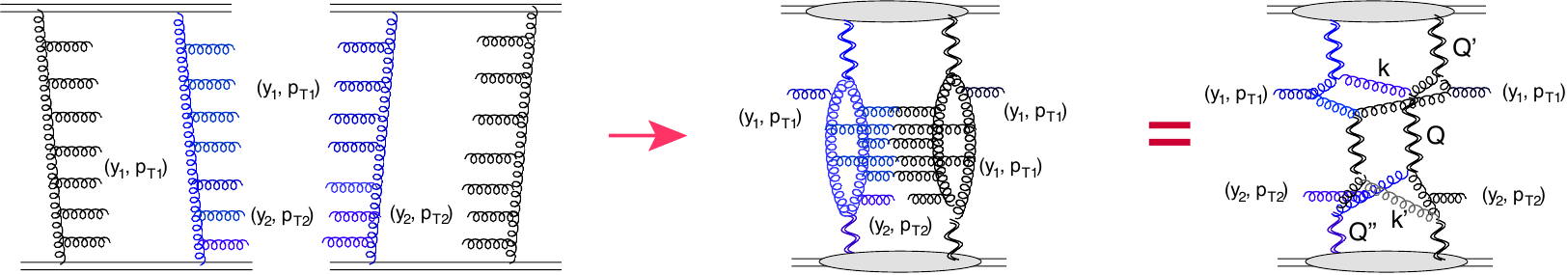}  
      \caption{ The  generalization of  \fig{becor}-b 
 for $\bas\,y_{12}\,\gg\,1$.
      The wavy lines show the BFKL Pomeron\cite{BFKL,LI}, while the
 helical lines denote  gluons.  The colour  (blue and black) of the lines indicates
 the parton shower. }
\label{mudia}
   \end{figure}

At large $y_{12}$, we have to take into account the emission gluons, 
  with rapidities $ y_2 \,<\,y_i\,<\,y_1$ which transform the Mueller
 diagram of \fig{becor}-b  to more general diagrams of \fig{mudia}. The
 general features of \fig{becor}-b  is that the lower Pomerons carry
 momenta $\vec{Q}_T + \vec{p}_{12}$ and $-\vec{Q}_T - \vec{p}_{12}
 $ with $\vec{p}_{12} = \vec{p}_{T1} - \vec{p}_{T2}$.  $\vec{Q}_T$
 denotes the momentum along the BFKL Pomeron. After integration over 
$Q_T$,
   we obtain $p_{12} \sim 1/R_h$, where $R_h$ is the
 size of the target(projectile), which has a non-perturbative origin. 
 Roughly speaking, the correlation function turns out to be proportional
 to $G\Lb p_{12}\Rb$, where $G$ denotes the non-perturbative form factor of 
the
 target  or projectile \cite{GOLE}.
This conclusion  stems from the value of the typical $Q_T$ for the BFKL
 Pomeron, which is determined by the size of the largest dipoles in the
 Pomeron. \fig{mudia} does not have these features. We will show that the
 azimuthal angle correlations  originate  from the integration over
 $\vec{Q}_T$(see \fig{mudia}), due to the structure of the vertices of
 emission of the gluons with $\vec{p}_{T1}$ and $\vec{p}_{T2}$, which
 have contributions proportional to $(\vec{p}_{T1}\cdot \vec{Q}_T)^n$
 $(\vec{p}_{T2}\cdot \vec{Q}_T)^n$. Recall, that these  kind of 
vertices,
 are the only possibilities to obtain  angular correlations in the
 classical Regge analysis\cite{KAID}. This mechanism for  azimuthal
 angular correlations was suggested in Ref.\cite{LERECOR} (see also
 Refs.\cite{MLSK,HHXY,GLMT,KOLUREV}), and in the 
review 
of Ref. \cite{KOLUREV}, was called, the density
 variation mechanism. 

The paper is organized as follows. In the next section we discuss
 the contribution of the diagram of \fig{mudia} in the momentum
 representation. In the  remainder of the paper, we will use the 
mixed
 representation: the dipole sizes and momentum transferred ($Q_T$),
 which will be introduced in section 3. Section 4 is devoted to the
 discussion of the single inclusive production in the Colour Glass
 Condensate (CGC)/saturation approach.  The double inclusive production
 is considered in section 5, in which the rapidity dependence of the master
 diagram of \fig{mudia} will be calculated.  
In section 6, we estimate the angular 
correlation
 function and Fourier harmonics $v_n$, and  we present our
 prediction for dependence of $v_n$ on the difference  of rapidities
 ($y_{12}$). In section 7 we draw our conclusions and outline 
 problems for future investigation.

 \section{Correlations in the momentum representation}
 The double inclusive cross section of \fig{mudia} takes the following form
 {\footnotesize\bea \label{MAEQ}
 &&\frac{d^2 \sigma}{d y_1 d^2 p_{T1}\,d y_2 d^2 p_{T2}}\Lb \fig{mudia}\Rb\,\,=\,\,\Lb\frac{2 C_F \as}{(2 \pi)^2 }\Rb^2\int \frac{d^2 k_T}{( 2 \pi )^2}\,\frac{d^2 k'_T}{( 2 \pi )^2} \,\frac{d^2 Q'_T}{( 2 \pi )^2}\,\frac{d^2 Q_T}{( 2 \pi )^2 }\,\frac{d^2 Q''_T}{( 2 \pi )^2 } \,k^2_T \,(\kv - \Qv)^2\\
  &&\times\,\,N\Lb Q'_T\Rb\, \phi^G_H\Lb -\vec{k}_T + \vec{Q}'_T, \vec{k}_T; Y - y_1\Rb\,\phi^G_H\Lb \vec{k}_T - \vec{Q}_T, -\vec{k}_T   + \vec{Q}_T - \vec{Q}'_T; Y -y_1\Rb\,\,\Gamma_\nu \Lb -\vec{k} + \vec{Q}_T, \vec{p}_{T1}\Rb \,\Gamma_\nu \Lb \vec{k} - \vec{Q}_T - \vec{Q}'_T, \vec{p}_{T1}\Rb\nn\\
  &&\times\,\, \phi\Lb-\vec{k}_T , -\vec{k}_T + \vec{Q}_T; \vec{k}'_t + \vec{p}_{T2},-\vec{k}'_T - \vec{p}_{2T} - \vec{Q}_T; y_{12}\Rb\,\phi\Lb- \vec{k}_T + \vec{Q}'_T  + \vec{p}_{T1}, \vec{k}_T - \vec{p}_{T1}  - \vec{Q}_T - \vec{Q}'_T; \vec{k}'_T - \vec{Q}''_T +\vec{Q}_T, \vec{k}'_T- \vec{p}_{T2}; y_{12}\Rb \nn\\
    &&\times\,\,N\Lb Q''_T\Rb\, \phi^G_H\Lb \vec{k}'_T - \vec{Q}''_T + \vec{Q}_T,- \vec{k}'_T - \vec{Q}_T; y_2\Rb\,\phi^G_H\Lb -\vec{k}'_T - \vec{Q}''_T   + \vec{Q}_T ,  \vec{k}'_T;y_2\Rb\,\Gamma_\mu \Lb -\vec{k}'_T  -\vec{p}_{T2} + \vec{Q}''_T ,\vec{p}_{T2}\Rb \,\Gamma_\mu \Lb \vec{k}'_T - \vec{p}_{T 2},\vec{p}_{T2}\Rb\nn
    \eea }
  
  where $\phi^G_H\Lb \vec{k}_T,-\vec{k} +\vec{Q}_T'\Rb $, as well as all 
other
 functions $\phi$ of this type,  are the correlation functions which at
 $Q'_T=0$, give the probability to find a gluon with  transverse 
momentum
 $\vec{k}_T$ in the hadron(nucleus) of the projectile (target).
  $\phi\Lb \vec{k}_T,-\vec{k} +\vec{Q}_T;  \vec{k}'_T,-\vec{k}'_T
 +\vec{Q}_T\Rb  $ describes the interaction of two gluons with momenta
 $\vec{k}_T$ and $\vec{k}'_T$,  which scatter at momentum transferred
 $Q"_T$. $N\Lb Q'_T\Rb$ is a pure phenomenological form factor that
 describes the probability to find  two Pomerons in the projectile
 or target, with transferred moment $\vec{Q}'_T$ and $-\vec{Q}'_T$.
 $C_F \,=\,\Lb N^2_c - 1\Rb/2 N_c$ where $N_c$ is the number of
 colours. The Lipatov vertex  $     \Gamma_\mu\Lb k_T,p_{T1}\Rb
  $ has the following form
  
    \beq \label{LV}
    \Gamma_\mu\Lb k_{T}, p_{T1}\Rb\,\,=\,\,\frac{1}{p_{12}^2}\Lb \,k^2_{T}\,
p_{T1,\mu}   \,-\,{k}_{T,\mu}\,p^2_{T1}\Rb;         \eeq  

Using \eq{LV} we obtain
\bea \label{KERN}
&&2\,\Gamma_\nu \Lb -\vec{k} + \vec{Q}_T, \vec{p}_{T1}\Rb \,\Gamma_\nu \Lb \vec{k} - \vec{Q}_T - \vec{Q}'_T, \vec{p}_{T1}\Rb\, =\nn\\
 && \frac{1}{p^2_{T1}}\Lb \Lb  -\vec{k} + \vec{Q}_T \Rb^2\,\Lb \vec{k}_T - \vec{p}_{T1} - \vec{Q}_T - \vec{Q}'_T\Rb^2\,+\,\Lb  -\vec{k} +\vec{p}_{T1} - \vec{Q}_T \Rb^2\,\Lb \vec{k}_T - \vec{Q}_T - \vec{Q}'_T\Rb ^2\Rb \,-\,Q'^2_T;\nn\\
 &&2\, \Gamma_\mu \Lb -\vec{k}'_T  -\vec{p}_{T2} + \vec{Q}''_T ,\vec{p}_{T2}\Rb \,\Gamma_\mu \Lb \vec{k}'_T - \vec{p}_{T 2} - \vec{Q}_T,\vec{p}_{T2}\Rb \,=\,\nn\\
 &&    \frac{1}{p^2_{T2}}  \Lb \Lb   -\vec{k}'_T  -\vec{p}_{T2} + \vec{Q}"_T \Rb^2  \Lb  \vec{k}'_T - \vec{Q}_T \Rb^2 \,+\,   \Lb   -\vec{k}'_T   + \vec{Q}''_T + \vec{Q}_T \Rb^2  \Lb  \vec{k} - \vec{p}_{T 2}- \vec{Q}''_T \Rb^2\Rb \,-\,Q''^2_T ;
     \eea
  We can simplify the master equation (see \eq{MAEQ} by 
observing,
 that dependence on $Q'_T$ and $Q''_T$   is determined  by
 the non-perturbative scale of the projectile(target) structure,
 which  in \eq{MAEQ}, is absorbed in the phenomenological form
 factors $N(Q'_T)$ and $N(Q''_T)$.  Therefore, the typical $Q'_T$
 and $Q''_T$ turn out to be of the order of the soft scale
 $\mu_{\rm soft}$, which is much smaller that the other typical
 momenta in \eq{MAEQ}, assuming that $P_{T1}$ and $P_{T2}$ are
 larger than $\mu_{\rm soft}$. Introducing 
  \beq \label{SOFTSC}
  \mu^2_{\rm soft}   \,\,=\,\,\int \frac{d^2 Q'_T}{(2\,\pi)^2} N\Lb Q'_T\Rb
  \eeq
  we can neglect $Q'_T$ and $Q''_T$ in the BFKL Pomeron Green's
  functions and  re- write \eq{MAEQ} in the form:
\bea \label{MAEQ1}
 &&\frac{d^2 \sigma}{d y_1 d^2 p_{T1}\,d y_2 d^2 p_{T2}}\Lb \fig{mudia}\Rb\,\,=\,\,\Lb\frac{2 C_F \as\,\mu^2_{\rm soft}}{(2 \pi)^2 }\Rb^2\int \frac{d^2 k_T}{( 2 \pi )^2}\,\frac{d^2 k'_T}{( 2 \pi )^2} \,\,\frac{d^2 Q_T}{( 2 \pi )^2 }\,  \,k^2_T \,(\kv - \Qv)^2
 \\
  &&\times\,\, \phi^G_H\Lb -\vec{k}_T , \vec{k}_T; Y - y_1\Rb\,\phi^G_H\Lb \vec{k}_T - \vec{Q}_T, -\vec{k}_T   + \vec{Q}_T;Y - y_1 \Rb\,\,\Gamma_\nu \Lb -\vec{k} + \vec{Q}_T, \vec{p}_{T1}\Rb \,\Gamma_\nu \Lb \vec{k} - \vec{Q}_T , \vec{p}_{T1}\Rb\nn\\
  &&\times\,\, \phi\Lb-\vec{k}_T , -\vec{k}_T + \vec{Q}_T; \vec{k}'_t + \vec{p}_{T2},-\vec{k}'_T - \vec{p}_{2T} - \vec{Q}_T;  y_{12}\Rb\,\phi\Lb- \vec{k}_T   + \vec{p}_{T1}, \vec{k}_T - \vec{p}_{T1}  - \vec{Q}_T ; \vec{k}'_T  +\vec{Q}_T, \vec{k}'_T- \vec{p}_{T2}; y_{12}\Rb \nn\\
    &&\times\,\, \phi^G_H\Lb \vec{k}'_T  + \vec{Q}_T,- \vec{k}'_T - \vec{Q}_T; y_2\Rb\,\phi^G_H\Lb -\vec{k}'_T   + \vec{Q}_T ,  \vec{k}'_T;y_2\Rb\,\Gamma_\mu \Lb -\vec{k}'_T  -\vec{p}_{T2}, \vec{p}_{T2}\Rb \,\Gamma_\mu \Lb \vec{k}'_T - \vec{p}_{T 2},\vec{p}_{T2}\Rb\nn
    \eea  
  with \eq{KERN} which takes the following form:
 {\small \bea \label{KERN1}
&&2\,\Gamma_\nu \Lb -\vec{k} + \vec{Q}_T, \vec{p}_{T1}\Rb \,\Gamma_\nu \Lb \vec{k} - \vec{Q}_T, \vec{p}_{T1}\Rb\, =\,\nn\\
&& \frac{1}{p^2_{T1}}\Lb \Lb  -\vec{k} + \vec{Q}_T \Rb^2\,\Lb \vec{k}_T - \vec{p}_{T1} - \vec{Q}_T \Rb^2\,+\,\Lb  -\vec{k} +\vec{p}_{T1} - \vec{Q}_T \Rb^2\,\Lb \vec{k}_T - \vec{Q}_T \Rb ^2\Rb \,-\,Q^2_T;\nn\\
 &&2\, \Gamma_\mu \Lb -\vec{k}'_T  -\vec{p}_{T2} ,\vec{p}_{T2}\Rb \,\Gamma_\mu \Lb \vec{k}'_T - \vec{p}_{T 2}- \vec{Q}_T,\vec{p}_{T2}\Rb \,=\,\nn\\
 &&  \frac{1}{p^2_{T2}}  \Lb \Lb   -\vec{k}'_T  -\vec{p}_{T2}  \Rb^2  \Lb  \vec{k}'_T- \vec{Q}_T \Rb^2 \,+\,   \Lb   -\vec{k}'_T    \Rb^2  \Lb  \vec{k} - \vec{p}_{T 2} - \vec{Q}_T \Rb^2\Rb \,-\,Q^2_T;
     \eea  }
     
     At high energies the parton densities $\phi(\dots; Y)$
 in \eq{MAEQ} and in \eq{MAEQ1}, are proportional to $\exp\Lb
 \Delta_{\rm BFKL} \,Y\Rb$ for the BFKL Pomeron, where
 $\Delta_{BFKL} = 2.8\, \bas$ is the intercept of the BFKL Pomeron.
  Bearing this in mind, one can see, that the interference diagram for
 the double inclusive cross section  does not depend on $y_1, y_2$ or on
 $y_{12}$.
     
The   main diagram of \fig{becor}-a, also does not depend on rapidities,
 and its expression  has the following form:

\bea \label{MAIN}
 &&\frac{d^2 \sigma}{d y_1 d^2 p_{T1}\,d y_2 d^2 p_{T2}}\Lb \fig{becor}-a\Rb\,\,=\,\,\Lb\frac{2 C_F \as}{(2 \pi)^2 } \int d^2 Q_T N^2\Lb Q_T\Rb \Rb^2\int \frac{d^2 k_T}{( 2 \pi )^2}\,\frac{d^2 k'_T}{( 2 \pi )^2} \nn\\
  &&\times\,\, \phi^G_H\Lb -\vec{k}_T , \vec{k}_T; Y - y_1\Rb\,\phi^G_H\Lb \vec{k}_T - \vec{p}_{T1}, - y_2 \Rb\,\,\Gamma_\nu \Lb -\vec{k} , \vec{p}_{T1}\Rb \,\Gamma_\nu \Lb \vec{k} , \vec{p}_{T1}\Rb\nn\\
  &&\times\,\, \ph^G_H\Lb-\vec{l}_T , \vec{l}_T ; Y -   y_{2}\Rb\,\ph^G_H\Lb \vec{l}_T   + \vec{p}_{T2}, -\vec{l}_T - \vec{p}_{T2}; y_2\Rb \,\Gamma_\mu \Lb \vec{l}_T , \vec{p}_{T2}\Rb \,\Gamma_\mu \Lb- \vec{l}_T ,\vec{p}_{T2}\Rb
    \eea

 
 \section{BFKL Pomeron in the mixed representation}
  For  a more convenient  presentation, it turns out that the most 
economical  way of
 calculating the diagram of \fig{mudia}, is to use the mixed
 representation of the BFKL Pomeron Green's function, $G\Lb
 \rv, \Rv,\Qv,Y\Rb$, where $r$ and $R$ are the sizes of two
 interacting dipoles , $Q_T$ denotes the momentum transferred by
 the Pomeron, and  $Y$  the rapidity between the two dipoles. 
  This Green's function is well known\cite{LI} and we discuss
 it here for the completeness of presentation, referring to
 Refs.\cite{LI,NAPE} for all details.  It has the following form:
  \beq \label{BFKLGF}
  G\Lb \rv, \Rv,\Qv; Y\Rb\,\,=\,\,\frac{r \,R}{16}\sum^{\infty}_{n=
 - \infty}\int^{\infty}_{-\infty} d \nu\frac{1}{\Lb \nu^2 +  
 \qu(n-1)^2\Rb\,\Lb \nu^2 + \qu(n+1)\Rb} V_{\nu,n}\Lb \rv,\Qv\Rb\,V^*_{\nu,n}\Lb\Rv,\Qv\Rb\,e^{ \omega\Lb \nu,n\Rb \,Y}
  \eeq
  where 
  \bea \label{OMEGA}
  \omega\Lb\nu,n\Rb \,\,&=&\,\,2\,\bas {\rm Re}\Lb \psi\Lb \h + 
\h |n| + \nu\Rb - \psi\Lb 1 \Rb \Rb;\nn\\
 \omega\Lb\nu,0\Rb \,\,&=&\,\,2\,\bas {\rm Re}\Lb \psi\Lb \h + 
+ \nu\Rb - \psi\Lb 1 \Rb \Rb \,\,\xrightarrow{\nu \ll 1}\,\,\Delta_{\rm BFKL}  \,\,-\,\,D\,\nu^2; \eea
  where $\psi(z)$  is the Euler $\psi$-function (see Ref.\cite{RY}
 formulae {\bf 8.36}) and   $\Delta_{\rm BFKL} =  \bas 4 \ln 2 ,   D = 
 \bas 14 \zeta(3), \xi = \ln\Lb r^2_1/r^2_2\Rb$.
   
  Each term in \eq{BFKLGF} has a very simple structure, being the
 typical contribution of the Regge pole exchange: the product of two
 vertices and Regge-pole propagator. From \eq{OMEGA} one can see
 that at large $Y$ the main contribution comes from the term with
 $n = 0,$ and in what follows we will concentrate on this particular  
term. 
   The vertices with $n=0$ have been  determined in 
Refs.\cite{LI,NAPE},
 and they have an elegant form in the complex number representation
 for the point on the two dimensional plane: viz.
   \beq \label{CN}
\mbox{For } \rv(x,y)  : \,\,   \rho = x + i y; \,\, \rho^*= x - i y;~~~~~
\mbox{For } \Qv(Q_x, Q_y)  : \,\,  q = Q_x + i Q_y;\,\, q^*= Q_x - i Q_y;
\eeq
  Using  this notation the vertices  
  have the following structure:
  \beq\label{V}
  V_{\nu}\Lb \rv,\Qv\Rb\,=\,\Lb Q^2_T\Rb^{i \nu}\,\Gamma^2\Lb 1 - i \nu\Rb\Bigg\{
  J_{-i \nu}\Lb \qu q^* \rho\Rb \,J_{- i \nu}\Lb \qu q \rho^*\Rb  \,\,-\,\, J_{i \nu}\Lb \qu q^* \rho\Rb \,J_{ i \nu}\Lb \qu q \rho^*\Rb \Bigg\}
     \eeq 
     
     At $Q_T \to 0$ this vertex takes the form:
   {\small  \bea \label{VSQ}
  && 2^{6 i  \nu}\,V_{\nu}\Lb \rv,\Qv\Rb\,\xrightarrow{Q_T r\,\ll\,1}  \\
 && \left(\frac{r^2}{2^6}\right)^{-i \nu } \left(\frac{(\nu +i) \Lb 8
 (\Qv\cdot \rv)^4 - 8(\Qv \cdot\rv)^2 Q^2_T  r^2 + 5\h Q^4_T r^4\Rb
 +(2 i +  \nu)Q^4_T r^4 )}{64^2 (\nu +2 i) (1-i \nu )^2}+\frac{i
 (\nu +i)\Lb(2 ( \Qv \cdot \rv)^2 - Q^2_T r^2\Rb}{32 (1-i \nu )^2}
   +1\right)\nn\\
   && + \left(Q^2\right)^{i \nu } \left(\frac{Q^2 r^2}{2^6}\right)^{i \nu }
 \left(\frac{(\nu -2 i)\Lb 8 (\Qv\cdot \rv)^4 - 8(\Qv \cdot\rv)^2 Q^2_T r^2 + 5\h Q^4_T r^4\Rb)}{2^{12} ((2+i \nu ) (1+i \nu ))^2}+\frac{2 (1+i \nu ) \Lb(2 ( \Qv \cdot \rv)^2 - Q^2_T r^2\Rb}{2^6 (1+i \nu )^2}-1\right)    \nn 
      \eea  }

     For small values of $\nu$ (which are related to the region of
 large $\bas Y \,\gg\,1$), \eq{VSQ} can be simplified and reduced to the 
form:
 \bea \label{VSQ1}
  &&2^{6 i \nu} V_{\nu}\Lb \rv,\Qv\Rb\,\xrightarrow{Q_T r\,\ll\,1}  \\
  && \left(\frac{r^2}{2^6}\right)^{-i \nu } \left(\frac{   (\Qv\cdot \rv)^4
 - (\Qv \cdot\rv)^2 Q^2_T  r^2 + \frac{9}{16} Q^4_T r^4  )}{2^8 }-\frac{2 ( \Qv \cdot \rv)^2 - Q^2_T r^2}{2^5 }
   +1\right)\nn\\
   && - \, \left(Q^2\right)^{i \nu } \left(\frac{Q^2 r^2}{2^6}\right)^{i
 \nu } \left(\frac{ (\Qv\cdot \rv)^4 - (\Qv \cdot\rv)^2 Q^2_T r^2 + \frac{9}{16} Q^4_T r^4)}{2^{8} }  -  \frac{   2 ( \Qv \cdot \rv)^2 - Q^2_T r^2}{2^5 } + 1\right)    \nn 
      \eea  
     Using  that 
     \beq \label{ASJ}
     J_{-i \nu}\Lb z \Rb\,\,\xrightarrow{z\,\gg\,1}\,\,\sin\Lb
 \qu \pi + z + \h  i \pi \nu\Rb \sqrt{\frac{2}{\pi}} \sqrt{\frac{1}{z}}
     \eeq
     at  $\nu \ll 1$ we obtain for $Q^2_T r^2 \gg 1$
      \beq \label{VLQ}
 V_{\nu}\Lb \rv,\Qv\Rb\,\,\xrightarrow{Q_T r\,\gg\,1} \,\, \Lb Q^2_T\Rb^{i \nu}\,
 \Gamma^2\Lb 1 - i \nu\Rb\,   \,\cos\Lb \h \Qv \cdot \rv\Rb\, 
 \frac{4 
 \,i\, \nu}{Q_T\, r}
 \eeq
     
 The contribution of the first term in \eq{BFKLGF} can be reduced
 to the following form for the scattering amplitude of two dipoles
 with sizes $r_1$ and $r_2$:
 \bea\label{N12}
 N\Lb r_1,r_2; Y\Rb\,\,&=&\,\,\frac{r_1\,r_2}{16}\int d \nu \frac{1}{\Lb \nu^2 + 1/4\Rb^2}\,V_\nu\Lb r_1,Q_t \to 0\Rb V^*_{\nu}\Lb r_2,Q_T \to 0\Rb\,e^{\omega\Lb \nu,0\Rb Y}\nn\\
 &=& \,\,\frac{r_1\,r_2}{16}\int d \nu \frac{1}{\Lb \nu^2 + 1/4\Rb^2}\,e^{\omega\Lb \nu,0\Rb Y} \Bigg\{(r^2_1)^{- i \nu} - \Lb \frac{Q^4_T r^2_1}{2^{12}}\Rb^{ i \nu}\Bigg\} \,\Bigg\{(r^2_2)^{ i \nu} - \Lb \frac{Q^4_T r^2_2}{2^{12}}\Rb^{ - i \nu}\Bigg\} \nn\\
 &=&\,\,\frac{r_1\,r_2}{16}\int d \nu \frac{1}{\Lb \nu^2 + 1/4\Rb^2}\,e^{\omega\Lb \nu,0\Rb Y} 2 \Lb\frac{r^2_2}{r^2_1} \Rb^{i \nu}\nn\\
 &\xrightarrow{Y \gg 1; \nu\ll 1}&\,\,2\, r_1\,r_2 \int d \nu  \exp\Lb  \Lb \bas \,4 \ln 2 \,- \bas 14 \zeta(3) \nu^2\Rb Y\Rb \Lb\frac{r^2_2}{r^2_1} \Rb^{i \nu} \nn\\
 &=& r_1\, r_2  \sqrt{\frac{2 \pi}{ D Y}}\exp\Lb \Delta_{\rm BFKL} Y - \frac{\xi^2}{4 D Y}\Rb
    \eea 
 where $ \Delta_{\rm BFKL}$ and $D$ are defined in \eq{OMEGA}.
    
    In  the derivation of \eq{N12} we neglected the contributions 
that
 are proportional to $\Lb \frac{Q^4_T r^2_2 \,r^2_1}{2^{12}}\Rb^{
 - i \nu} $ since this contribution will be the same as in \eq{N12},
 but with, $\xi = \ln  \Lb \frac{Q^4_T r^2_2 \,r^2_1}{2^{12}}\Rb 
\,\gg\,1$.
  To integrate over $\nu$, we use the method of steepest descent, and 
the expansion of $\omega\Lb \nu,0\Rb$ at small $\nu$ (diffusion approximation, see the second equation in \eq{OMEGA}).
 
 $ N\Lb r_1,r_2; Y\Rb$ denotes the imaginary part of the dipole-dipole
 sacttering amplitude at $Q_T=0$, which is related to the cross section.
  One can check that \eq{N12} has the correct dimension.
 
 
 \section{Single inclusive production in a one parton shower}
 
 \subsection{BFKL Pomeron: the simplest approach for  a one parton shower}
 
 The single inclusive cross section  resulting from the one BFKL 
Pomeron is known,
 and it is equal to
 {\small\beq \label{SINCL}
 \frac{d^2 \sigma}{d y d^2 p_{T}}\,\,=\,\,\frac{2 C_F \as}{(2 \pi)^2}\int
 \frac{d^2 d^2 k_T}{(2 \pi)^2} \phi^G_H\Lb \vec{k}_T, Q_T=0; Y -
 y\Rb \phi^G_H\Lb \vec{k}_T - \vec{p}_T, Q_T=0; y \Rb \,\Gamma_\nu\Lb \vec{k}_T,\vec{p}_T\Rb\,\Gamma_\nu\Lb -\vec{k}_T,\vec{p}_T \Rb
 \eeq}
 
 The relation between the parton densities $\phi$ and the Green's
 function of the BFKL Pomeron has been  given in Ref.\cite{AGK2}:
 \beq \label{SINCL1}
 N^{\rm \tiny BFKL}\Lb r, r_1; y, Q_T=0\Rb 
\,\,=\,\,\frac{\as}{2} \int d^2 k_T \,\Lb 1\,-\,e^{i \vec{k}_T \cdot \vec{r}}\Rb\, \frac{\phi^G_H\Lb \vec{k}_T, Q_T=0; y \Rb}{k^2_T} \eeq
 where $ N^{\rm \tiny BFKL}\Lb r, r_1; Y\Rb $ is given by  \eq{BFKLGF}
 or by \eq{N12}, in the high energy limit. 
 \eq{SINCL1} can be re-written as follow
 \beq \label{SINCL11}
  \phi^G_H\Lb \vec{k}_T , Q_T=0; y \Rb \,\,=\,\,\frac{2}{\as}\int d^2 e^{i \vec{k}_T \cdot \vec{r}}\,\nabla^2_{r} \,
  N^{\rm \tiny BFKL}\Lb r, r_1; y, Q_T=0\Rb 
  \eeq 
  We have 
 \beq \label{SINCL2}
 \Gamma_\nu\Lb \vec{k}_T,\vec{p}_T\Rb\,\Gamma_\nu\Lb -\vec{k}_T,\vec{p}_T\Rb  \,\,=\,\,\frac{k^2_T\,\Lb \vec{k}_T - \vec{p}_T\Rb^2}{p^2_T}
 \eeq
 Plugging in \eq{SINCL1} and \eq{SINCL2} into \eq{SINCL} we obtain
\cite{AGK2}
 \beq \label{SINCL3}
  \frac{d^2 \sigma}{d y\, d^2 p_{T}}\,\,=\,\,\frac{8 C_F }{\as\,(2 \pi)^2} \frac{1}{p^2_T}\int d^2 r \,e^{i \vec{p}_T \cdot \vec{r}}\,\nabla^2 _r\,N^{\rm \tiny BFKL}_{\rm pr}\,\Lb r, r_1; Y- y, Q_T=0\Rb \,\nabla^2 _r\,N^{\rm \tiny BFKL}_{\rm tr}\,\Lb r, r_2; y, Q_T=0\Rb  
  \eeq 
  where $N_{\rm pr}$ and $N_{\rm tr}$  denote the probability to
 find a dipole in the projectile and target, respectively. $r_1$ and
 $r_2$ are the typical dipoles sizes in the projectile and target.
  
  As  can be seen from \eq{MAEQ} we need to generalize \eq{SINCL3} for
 the case $Q_T \neq 0$. \eq{SINCL} has to be replaced by

   \bea \label{SINCL4}
&& \frac{d^2 \sigma}{d y d^2 p_{T}}\Lb Q_T \neq 0\Rb \,\,=\\
&&\frac{2 C_F \as}{(2 \pi)^2}\int \frac{ d^2 k_T}{(2 \pi)^2} \phi^G_H\Lb \vec{k}_T, Q_T, Y - y\Rb \phi^G_H\Lb \vec{k}_T - \vec{p}_T, Q_T; y \Rb \,\Gamma_\nu\Lb \vec{k}_T,\vec{p}_T\Rb\,\Gamma_\nu\Lb -\vec{k}_T + \vec{Q}_T,\vec{p}_T \Rb\nn
 \eea 
  
Taking into account \eq{SINCL1} for $Q_T \neq 0$ and 
{\small\beq \label{SINCL5}
  \Gamma_\nu\Lb \vec{k}_T,\vec{p}_T\Rb\,\Gamma_\nu\Lb -\vec{k}_T + \vec{Q}_T,\vec{p}_T \Rb\,\,=\,\,\h\Bigg\{ \frac{1}{p^2_T}\Bigg[ 
  \Lb \vec{k}_T - \vec{Q}_T\Rb^2 \Lb \vec{k}_T - \vec{p}_T\Rb^2 \,\,+\,\, \Lb \vec{k}_T\Rb^2 \Lb \vec{k}_T - \vec{p}_T - \vec{Q}_T\Rb^2\Bigg]\,\,-\,\,Q^2_T\Bigg\}
 \eeq}
  we re-write \eq{SINCL3} in the form
  \bea \label{SINCL6}
    \frac{d^2 \sigma}{d y\, d^2 p_{T}}\Lb Q_T \neq 0\Rb\,\,&=&\,\, \frac{4 C_F }{\as\,(2 \pi)^2} \frac{1}{p^2_T}\int d^2 r \,e^{i \vec{p}_T \cdot \vec{r}}\\
  &\times&\Bigg\{-\nabla^2 _r\,N^{\rm \tiny BFKL}_{\rm pr}\,\Lb r, r_1; Y - y, Q_T\Rb \,( - i \nabla_r\, -\,\vec{Q}_T)^2\,N^{\rm \tiny BFKL}_{\rm tr}\,\Lb r, r_2; y, Q_T\Rb  \,\nn\\
  &+&\,( - i \nabla_r\, -\,\vec{Q}_T)^2\,(-\nabla^2_r)N^{\rm \tiny BFKL}_{\rm pr}\,\Lb r, r_1; Y  - y, Q_T\Rb \,\,N^{\rm \tiny BFKL}_{\rm tr}\,\Lb r, r_2;  y, Q_T\Rb\Bigg\}\nn\\
 & -& \,\, Q^2_T   \frac{4 C_F }{\as\,(2 \pi)^2}  \int d^2 r \,e^{i \vec{p}_T \cdot \vec{r}}\,N^{\rm \tiny BFKL}_{\rm pr}\,\Lb r, r_1; Y - y, Q_T\Rb \,N^{\rm \tiny BFKL}_{\rm tr}\,\Lb r, r_2; y, Q_T\Rb\nn   \eea   
 
  \subsection{General estimates}
 It should be stressed that the single inclusive production has the
 form of \eq{SINCL3} and \eq{SINCL6}  as  was shown in Ref.\cite{AGK2}
 for the general structure of the single parton shower.  For example,
 for the  process  shown in \fig{becor}-c. We need  only  to 
substitute    $N^{G}_{\rm tr}\,\Lb r, r_2;  y, Q_T\Rb$ for
  $2 N^{\rm \tiny BFKL}_{\rm tr}\,\Lb r, r_2; 
 y, Q_T\Rb$   where
{\small \beq \label{SINCL7}
  2 N^{\rm \tiny BFKL}_{\rm tr}\,\Lb r, r_2;  y, Q_T\Rb\,\,\to\,\, N^{G}_{\rm tr}\,\Lb r, r_2;  y, Q_T\Rb\,\,
 =\,\,2 N_{\rm tr}\,\Lb r, r_2;  y, Q_T\Rb  
  -  \int d^2 Q'_T\, N_{\rm tr}\,\Lb r, r_2;  y, \vec{Q}_T - \vec{Q}'_T\Rb\,  N_{\rm tr}\,\Lb r, r_2;  y,  \vec{Q}'_T\Rb
  \eeq}
  
   $N_{\rm tr}\,\Lb r, r_2;  y, Q_T\Rb$ is a solution to the
 non-linear evolution equation.  For the case of  inclusive
 production, we can considerably simplify estimates noting that 
  \beq \label{SINCL8}
   \nabla^2_r \,N_{\rm tr}\,\Lb r, r_2;  y, Q_T\Rb  \xrightarrow{r^2 Q^2
_s\Lb y\Rb \,\ll\,1}\,\,\, N^{\rm BFKL}_{\rm tr}\,\Lb r, r_2;  y, Q_T\Rb\,\,\ll\,\,1;   
  ~~~~    
   \nabla^2_r \,N_{\rm tr}\,\Lb r, r_2;  y, Q_T\Rb
  \xrightarrow{r^2 Q^2_s\Lb y\Rb\, \gg \,1}\,\,0;
    \eeq
   where $Q_s(y)$ denotes the saturation momentum.
   
  In other words, the main contribution to  inclusive production
 comes from the vicinity of the saturation scale, where $r^2 Q^2_s 
\approx\,1$. Fortunately, the behavior of $N$ in this kinematic region
 is determined by the linear BFKL evolution equation \cite{GLR,MUQI,MV}
 and has the following form\cite{MUTR}
  \beq \label{VICQS}
  N_{\rm tr}\,\Lb r, r_2;  y, Q_T = 0\Rb \,\,\propto\,\,\Lb r^2 Q^2_s(y)\Rb^{1 - \gamma_{cr}}~\mbox{with}~~ Q^2_s =(1/r^2_2)\,\exp\Lb \frac{\omega\Lb\gamma = \h + i \nu = \gamma_{cr}\Rb}{1 - \gamma_{cr}}\,y\Rb\,=\,(1/r^2_2)e^{\kappa\,y}
  \eeq
  where $ \gamma_{cr}$ = 0.37.
  
 We have seen in \eq{VLQ} that for $Q_T \neq 0$
,  the scattering amplitude decreases
 at $Q^4_T\, r^2 \,r^2_2\, \gg \,1$. Therefore, we need to consider
 rather small values of $Q_T $: $Q^4_T\, r^2 \,r^2_2\, \leq  \,1  $.
 The product of vertices that determines the amplitude has two terms
( see \eq{VSQ} )  which are proportional to  $\Lb r^2/r^2_2\Rb^{i \nu}$
 and  to $\Lb Q^4_T\, r^2 \,r^2_2\Rb^{I \nu}$. Therefore,  
  the maximum of $\nabla^2_r N$  can be reached if  $r^2/r^2_2 
 e^{\kappa\,y} \,\sim\, 1$ and $ Q^4_T\,r^2\,r^2_2 e^{\kappa\,y}
\,\sim\,1$ and the amplitude  then has the following form
 \beq \label{VICQS1}     
    N_{\rm tr}\,\Lb r, r_2;  y, Q_T\Rb \,\,\propto\,\, c_1 \Lb \frac{r^2}{r^2_2} e^{\kappa\,y}\Rb^{1 - \gamma_{cr}}\,\,+\,\,c_2 \Lb   Q^4_T\, r^2 \,r^2_2\,e^{\kappa\,y}\Rb^{1 - \gamma_{cr}}
    \eeq
     
    The first term does not depend on $Q_T$ and, therefore, 
 the upper limit of the integral
 over $ Q_T$, goes up to $\Lb Q^{max}_T\Rb^2\,\approx 1/(r\,r_2)$. The second term both for $Q^2_T \,r\,r_2\,<\,e^{- \h \kappa \,y}$ and for $Q^2_T \,r\,r_2\,>\,e^{- \h \kappa \,y}$ turns out to be small.   Indeed,  in the first region the amplitude is
 small,  while in  the second region we are deep in the saturation domain 
where
 $\nabla^2_r N \to 0$.  Hence, we expect that in the integral over $Q_T$,
 the first term gives a larger contribution  than the second term and we will keep only this contribution in our estimates.
  \section{Double inclusive cross section  for two parton shower 
production}
  
  \subsection{The simplest diagram}
     
     In this section we calculate the simplest diagram of \fig{mudia}.
  We need to integrate  the
 product of two BFKL Pomerons over $Q_T$ (see \eq{MAEQ1}):
     \beq \label{DISD1}
 I\,\,=\,\,    \int d^2 Q_T\,V_{\nu_1}\Lb \vec{r}_1,\vec{Q}_T\Rb\,V^*_{\nu_1}\Lb \vec{r}_2,\vec{Q}_T\Rb \,   V_{\nu_2}\Lb \vec{r}'_1,\vec{Q}_T\Rb\,V^*_{\nu_2}\Lb \vec{r}'_2,\vec{Q}_T\Rb 
     \eeq
   From \eq{MAEQ1} in the momentum representation, we see that $r_1\, 
\neq\, r'_1$($  r_2\, \neq\, r'_2$  ) but they are close to each other,
 being determined by the same momentum $k_T$. We assume that $p_{T1}
 < k_T$, since 
   $k_T \sim Q_s\Lb Y - y_1\Rb \gg \mu_{\rm soft}$. Considering  $r_1 
\approx r'_1 \,\ll\,r_2 \approx r'_2$ we will show that in the integral
 over $Q_T$, the typical $Q_T \sim 1/r_2$. In other words, the dependence
 of $Q_T$ is determined by the largest of interacting dipoles.
   
   From \eq{VLQ} we see that for large $Q_T$, when $r^2_1 Q^2_T \,\gg\,1$
 and $r^2_2\,Q^2_T\,\gg\,1$, the integrand is proportional to $1/Q^4_T$,
 and converges.  The main  region  of interest is  $r^2_2\, Q^2_T 
\,\gg\,1$
 and  $r^2_1\,Q^2_T\,\ll\,1$. In this kinematic region for vertices
 $V_{\nu_1}\Lb \vec{r}_1,\vec{Q}_T\Rb$\, and   
  $ V_{\nu_2}\Lb \vec{r}'_1,\vec{Q}_T\Rb$, we can use \eq{VSQ1}, while
 the conjugated vertices are still in the regime of \eq{VLQ}.  
 \eq{DISD1}  then takes the form:
  \bea \label{DISD2}
&&   I\,=\,  2^{6  i (\nu_1 + \nu_2)} \Lb - 16\nu_1\,\nu_2\Rb\pi  \\
&&\times\int_{1/r^2_2}\,\, d Q^2_T\,\Bigg\{ \Lb \frac{Q^2\,r^2_1}{2^6}\Rb^{-i  \nu_1}\,-\,\Lb \frac{Q^2\,r^2_1}{2^6}\Rb^{i  \nu_1}\Bigg\}  
  \Bigg\{ \Lb \frac{Q^2\,r'^2_1}{2^6}\Rb^{-i  \nu_2}\,-\,\Lb \frac{Q^2\,r'^2_1}{2^6}\Rb^{i  \nu_2}\Bigg\} \frac{\cos^2\Lb \h \Qv \cdot \vec{r}_1\Rb}{Q^2_T \,r^2_2}\nn
  \eea
   Assuming that both $\nu_1$ and $\nu_2$ are small, we see that
 all four terms are equal to each other, and the integral can be
  written as follows.
     \beq \label{DISD3}   
  I\,\,=\,\, 2^{6  i (\nu_1 + \nu_2)} \Lb - 2^6\nu_1\,\nu_2\Rb\pi \frac{1}{i\,(\nu_1 + \nu_2)}  \Lb \frac{r^2_1}{r^2_2}\Rb^{i\,(\nu_1 + \nu_2)}\frac{1}{r^2_2}
  \eeq
  
  The appearance of the pole $\nu_1=-\nu_2$ indicates that the contribution
 from this kinematic region is large.
  
  Closing the contour of integration on $ \nu_2$
  over the pole, we obtain 
  \beq \label{DISD4}
  I\,\,=\,\,2^6 \pi\,\nu^2_1\frac{1}{r^2_2}
  \eeq  
  
  Actually,   the double inclusive cross section depends on $\nabla
  ^2 N$ as we argued in the previous section. Repeating the procedure for 
     \bea \label{DISD5}
&& {\cal I}\,\,=\,\,  \\
 &&  \int d^2 Q_T\,\nabla^2_{r_1}\Lb r_1\,V_{\nu_1}\Lb \vec{r}_1,\vec{Q}_T\Rb\Rb\,\nabla^2_{r'_1}\Lb r'_1\,V^*_{\nu_1}\Lb \vec{r}_2,\vec{Q}_T\Rb\Rb \,  \nabla^2_{r_2} \Lb r_2\,V_{\nu_2}\Lb \vec{r}'_1,\vec{Q}_T\Rb\Rb\,\nabla^2_{r'_2}\Lb r'_2\,V^*_{\nu_2}\Lb \vec{r}'_2,\vec{Q}_T\Rb \Rb\nn
     \eea  
  we obtain for small $\nu_1$ and $\nu_2$:
  \beq \label{DISD6}
 {\cal I}\,\,=\,\,2^6 \pi\,\nu^2_1\frac{1}{r_1\,r'_1\,r^2_2\,r'^2_2}
  \eeq

  Taking the integral over $\nu_1$, using the method of steepest descent, 
we
  obtain the following contribution:
    \beq \label{DISD7}
 {\cal I}\,\,=\,\,2^5 \,\frac{1}{r_1\,r'_1\,r^2_2\,r'^2_2}\sqrt{\frac{\pi}{\Lb 2 \,D\,y_{12}\Rb^3}}\,\,e^{ 2 \Delta_{\rm BFKL} \,y_{12}}
  \eeq   
 where $ \Delta_{\rm BFKL}$ and $D$ are defined in \eq{OMEGA}.

Rewriting   \eq{MAEQ1} in the coordinate representation we obtain:
\bea \label{MAEQC1}
 &&\frac{d^2 \sigma}{d y_1 d^2 p_{T1}\,d y_2 d^2 p_{T2}}\Lb \fig{mudia}\Rb\,\,=\,\,\Lb\frac{2 C_F \,\mu^2_{\rm soft}}{\as\,(2 \pi)^2 }\Rb^2\,\,\frac{1}{p^2_{T1}\,p^2_{T2}}\,\int \frac{d^2 Q_T}{(2\,\pi)^2}\\
 &&\times\,\,\int  d^2 r_1\,d^2  r'_1\, d^2 \tilde{r} _1\,d^2 \tilde{r}'_1  \,e^{-i\, \vec{p}_{T1} \cdot \vec{\tilde{r}}'_1}\delta^{(2)}\Lb \vec{r}_1 + \vec{r}'_1 - \vec{\tilde{r}}_1 - \vec{\tilde{r}}'_1\Rb  \,\nabla^2_{r_1} N_{\rm pr}\Lb r_1; Y - y_1\Rb\,e^{i \vec{Q}_T \cdot \vec{\tilde{r}'}_1}\,\nabla^2_{\tilde{r}'_1} N_{\rm pr}\Lb\tilde{r}'_1; Y - y_1\Rb \nn\\
&&\times\,\,
\nabla^2_{\tilde{r}_1} \nabla^2_{\tilde{r}_2} N\Lb \tilde{r}_1; \tilde{r}_2, Q_T; y _{12}\Rb    \nabla^2_{\tilde{r}'_1} \nabla^2_{\tilde{r}'_2} N\Lb \tilde{r}'_1; \tilde{r}'_2,  Q_T; y _{12}\Rb \nn\\ 
&&\times\,\,\int \, d^2 r_2\,d^2 r'_2\, d^2 \tilde{r} _2\,d^2 \tilde{r}'_2\,e^{- i\,\vec{p}_{T2} \cdot \vec{\tilde{r}}_2}   \,\delta^{(2)}\Lb \vec{r}_2 + \vec{r}'_2 - \vec{\tilde{r}}_2 - \vec{\tilde{r}}'_2\Rb\,\nabla^2_{r_2} N_{\rm tr}\Lb r_2; y_2\Rb\,e^{i \vec{Q}_T \cdot \vec{\tilde{r}'}_2}\,\nabla^2_{\tilde{r}'_2} N_{\rm tr}\Lb\tilde{r}'_2; y_2\Rb\nn\eea

In \eq{MAEQC1} we neglected the terms which are proportional to
 $Q^2_T$ (see \eq{MAEQ1})  as the typical $Q_T$ are
 small as we have argued, and because  these terms do not lead
 to additional correlations in the azimuthal angles.

We have discussed the integral over $Q_T$, and 
it has the form of
 \eq{DISD7}.  The extra $e^{i \vec{r}'_1 \cdot \vec{Q}_T}$
 give an additional numerical factor, replacing $2^5$ by $2^7$ in
 \eq{DISD7}. To  integrate over $k_T$ and $k'_T$ we replace 
 \beq \label{DISD8}
 \int \prod d \phi_i\,e^{-i\, \vec{p}_{T1} \cdot \vec{\tilde{r}}'_1}\delta^{(2)}\Lb \vec{r}_1 + \vec{r}'_1 - \vec{\tilde{r}}_1 - \vec{\tilde{r}}'_1\Rb \,\to\, \Lb 2 \pi\Rb^4 \int k_T d k_T\,J_0\Lb k_T r_1\Rb \,J_0\Lb |\vec{k}_T + \vec{p}_{T1}|\,\tilde{r}'_1\Rb \,J_0\Lb k_T \tilde{r}_1\Rb \,J_0\Lb k_T \tilde{r}_1\Rb 
 \eeq
 Now we can take the integrals over $r_i$ bearing in mind \eq{DISD7} and 
 \beq \label{DISD9}
 N_{pr}\Lb r_i, Y - y_1\Rb= \int \frac{d \nu}{2 \pi} \Lb \mu^2_{\rm soft}\,r^2_i\Rb^{\h + i \nu_i}\,e^{\omega\Lb \nu_i,0\Rb \Lb Y - y_1\Rb}
 \eeq
 The integrals over $\tilde{r}_1 $and $\tilde{r}'_1$ have the
 following form (see Ref.\cite{RY} formulae {\bf 6.511(6)})

 \begin{displaymath} \label{DISD10} 
  \int^{\tilde{r}_2}_0 J_0\Lb k_T \tilde{r}_1\Rb \,d \tilde{r}_1\,\,=\,\, \tilde{r}_2\,J_0\Lb k  \tilde{r}_2\Rb + \h \pi \tilde{r}_2 \Big(
  J_1\Lb  k  \tilde{r}_2\Rb \,{\bf H}_0\Lb  k  \tilde{r}_2\Rb\,-\,  
 J_0\Lb  k  \tilde{r}_2\Rb \,{\bf H}_1\Lb  k  \tilde{r}_2\Rb  \Big)
 = \left\{ \begin{array}{ll}
 \tilde{r}_2 &\textrm{if $k  \tilde{r}_2\,\ll\, 1$}\\
\frac{1}{k} & \textrm{if $k  \tilde{r}_2\,\gg\, 1$}\\
\end{array} \right.
\end{displaymath}  
 Using \eq{DISD9}  we obtain 
  \beq \label{DISD11}
    \int^{\infty}_0\,r_i\,d r_i\,J_0\Lb k_T \,r_i\Rb,\nabla^2_{r_i} N_{pr}\Lb r_i, Y - y_1\Rb   \,\,=\,\, \frac{1}{k} \Lb \frac{4 \mu^2_{\rm soft}}{k^2}\Rb^{i \,\nu_i}
    \,e^{\omega\Lb \nu_i,0\Rb\, \Lb Y - y_1\Rb}
\eeq
Collecting \eq{DISD9},\eq{DISD10} and \eq{DISD11} we see that
 the main contribution stems from the region $ k  \tilde{r}_2\,\ll\, 1$
 and the integral over $k_T$ has the form
  \bea \label{DISD12}
   &&\tilde{r}_2 \tilde{r'}_2   \int^{1/\tilde{r}_2^2}_{p^2_{T1}} \frac{d k^2_T}{k^2_T} \Lb \frac{4 \mu^2_{\rm soft}}{k^2}\Rb^{i \,(\nu_1 +\nu'_1)}\,e^{(\omega\Lb \nu_i, 0\Rb\,+\omega\Lb 0,\nu'_1\Rb) \Lb Y - y_1\Rb} \,\,=\\
   && \tilde{r}_2 \tilde{r'}_2\frac{1}{i \,(\nu_1 +\nu'_1)}   
 \Lb  \frac{1}{ \tilde{r}_2^2 \, p^2_{T1}}\Rb^{i \,(\nu_1 +\nu'_1)} e^{(\omega\Lb \nu_i, 0\Rb\,+\omega\Lb 0,\nu'_1\Rb) \Lb Y - y_1\Rb} \,\nn\\
          &&\xrightarrow{\mbox{after integration over } \nu_1, \nu_2}\,\,\h\,  \sqrt{\frac{\pi}{2 \,D\,(Y - y_1)}}\,e^{ 2 \Delta_{\rm BFKL} \Lb Y - y_1\Rb }\nn
               \eea

   The integral over $k'_T$ has the same structure  while the
 integration in \eq{DISD10}  goes to infinity. As the result we 
can reduce  the integral to the form;
   
   \bea \label{DISD13}
  && \int_{p^2_{T2}} \frac{d k^2_T}{k^4_T} \Lb \frac{4 \mu^2_{\rm soft}}{k^2}\Rb^{i \,(\nu_1 +\nu'_1)}\,e^{(\omega\Lb \nu_i, 0\Rb\,+\omega\Lb 0,\nu'_1\Rb) \Lb Y - y_1\Rb} \,\,\\
  &&=\,\,\frac{1}{1 + i (\nu_1+\nu'_1)}\frac{1}{4 \mu^2_{\rm soft}}\Lb \frac{4 \mu^2_{\rm soft}}{p^2_{T2}}\Rb^{1+i \,(\nu_1 +\nu'_1)}\,e^{(\omega\Lb \nu_i, 0\Rb\,+\omega\Lb 0,\nu'_1\Rb) \Lb Y - y_1\Rb}\nn\\
   &&\xrightarrow{\mbox{after integration over } \nu_1, \nu'_1} \,\,\, \h \frac{\pi}{\,D\,y_2}\,\frac{1}{p^2_{T2}}\nn
   \eea

   Finally,
      \beq \label{DISD14}   
     \frac{d^2 \sigma}{d y_1 d^2 p_{T1}\,d y_2 d^2 p_{T2}}\Lb \fig{mudia}\Rb\,\,=
     \,\,4 \pi \Lb\frac{2 C_F}{\as\,(2 \pi)^2 }\Rb^2\,\,\frac{1}{p^2_{T1}\,p^4_{T2}} \, \sqrt{\frac{1}{2 \,D\,(Y - y_1)}}\sqrt{\frac{1}{\Lb 2 \,D\,y_{12}\Rb^3}}\, \frac{1}{\,D\,y_2}\,e^{ 2 \Delta_{\rm BFKL }\,Y}
     \eeq
  \subsection{The CGC/saturation approach}
  The integral over $k'_T$ in \eq{DISD13}  has an infra-red singularity
 with a cutoff  at $p_{T2}$ since we  assume that
 $p_{T2}$ is the smallest momentum. This reflects the principle feature
 of the BFKL Pomeron parton cascade, which has diffusion, both in the
 region of small and large transverse momenta. On the other hand we know
 that the CGC/saturation approach suppressed the diffusion in the small
 momenta\cite{KOLEB}, providing the natural  cutoff  for the 
infrared divergency. We expect that such a cutoff will be the value of the smallest 
saturation momenta: $ Q_s\Lb Y - y_1\Rb$ or $Q_s\Lb y_2\Rb$, which will replace one of $p^2_{T2}$ in the dominator of \eq{DISD14}.  Therefore, we anticipate that for the  realistic structure of the one parton
 shower cascade, (see \fig{becor}-c for example),  the contribution for 
the
 double inclusive cross section will be different.

     We need to specify the behavior of the scattering amplitude in the
 vicinity of the saturation scale. We have discussed the basic
 formulae\cite{MUTR}  of \eq{VICQS},  but for integration over the dipole 
sizes we need to know 
     the size of this region.    The scattering amplitude can be written
 in the form:
     \beq \label{AM1}
    N\Lb r_1,r_2; Y\Rb\,\,=\,\,\int^{\epsilon + i \infty}_{\epsilon - i \infty}\frac{d \gamma}{2 \pi }\,n_{in}\Lb \gamma\Rb
    e^{ \omega\Lb \gamma, 0\Rb Y - ( 1 - \gamma) \xi}
    \eeq
    where $ \omega(\gamma,0)$ is given by \eq{OMEGA},  replacing $\h + i 
\nu \equiv \gamma$ and $\xi = \ln\Lb r^2_1/r^2_2\Rb$. The saturation 
scale is deterring by the line on which the amplitude is a constant 
 (C) of the order one. This leads to the following equation for the
 saturation scale\cite{GLR,MUTR}:
    \beq \label{QS}
    \omega\Lb \gamma_{cr}, 0\Rb Y - ( 1 - \gamma_{cr}) \xi_s\,=\,0;~~~~   \omega'_\gamma\Lb \gamma, 0\Rb Y -  \xi_s\,=\,0;
    \eeq        
    which results in the value of $\gamma_{cr}$ given by the equation:
    \beq \label{GACR}
   \frac{ \omega\Lb \gamma_{cr}, 0\Rb }{1 - \gamma_{cr}}\,\,=\,\,  \omega'_\gamma\Lb \gamma, 0\Rb
   \eeq
   which gives $\gamma_{cr}= 0.37$ and the equation for the saturation
 momentum:
   \beq \label{QS1}
   \xi_s\,\,\equiv\,\,\ln\Lb Q^2_s\,r^2_2\Rb\,=\,\kappa Y\,=\,\frac{\omega\Lb \gamma_{cr}\Rb }{1 - \gamma_{cr}}\,Y
   \eeq
   
   Expanding the phase $   \omega\Lb \gamma, 0\Rb Y - ( 1 - \gamma) \xi$ in the vicinity $\Delta \xi = \xi - \xi_s   $ and $\Delta \gamma =  \gamma - \gamma_{cr}$  we obtain
    {\small \beq \label{AM2}
    N\Lb r_1,r_2; Y\Rb\,\,=\,\,\mbox{C}\,\int^{\epsilon + i \infty}_{\epsilon - i \infty}\frac{d \gamma}{2 \pi }\,\Lb r^2_1\,Q^2_s\Rb^{1 - \gamma_{cr}}\int\frac{d \Delta \gamma}{2 \pi i}
    e^{ \h \omega''_{\gamma \gamma}\Lb \gamma, 0\Rb Y(\Delta \gamma)^2 +  \Delta \gamma \Delta \xi }\,
    =\,  \Lb r^2_1\,Q^2_s\Rb^{1 - \gamma_{cr}}\mbox{C}\sqrt{\frac{\pi}{D Y}}\,e^{ - \frac{(\Delta \xi)^2}{4 D Y}}
     \eeq  }
    
\begin{figure}[ht]
    \centering
  \leavevmode
  \begin{tabular}{ccc}
      \includegraphics[width=6cm]{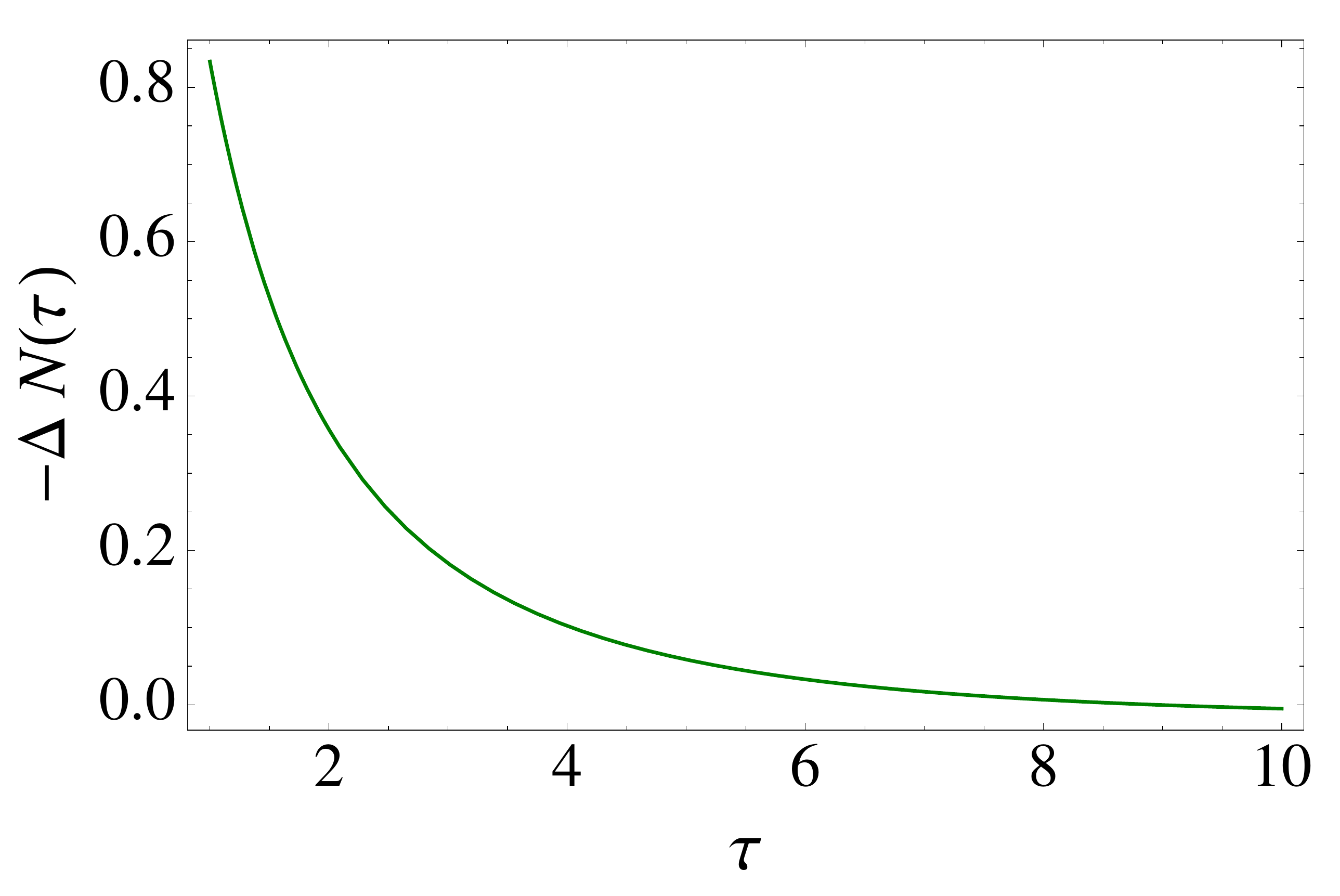} &~~~~~~~~~~~~~~~~~~~~~~~&\includegraphics[width=4cm,height=4cm]{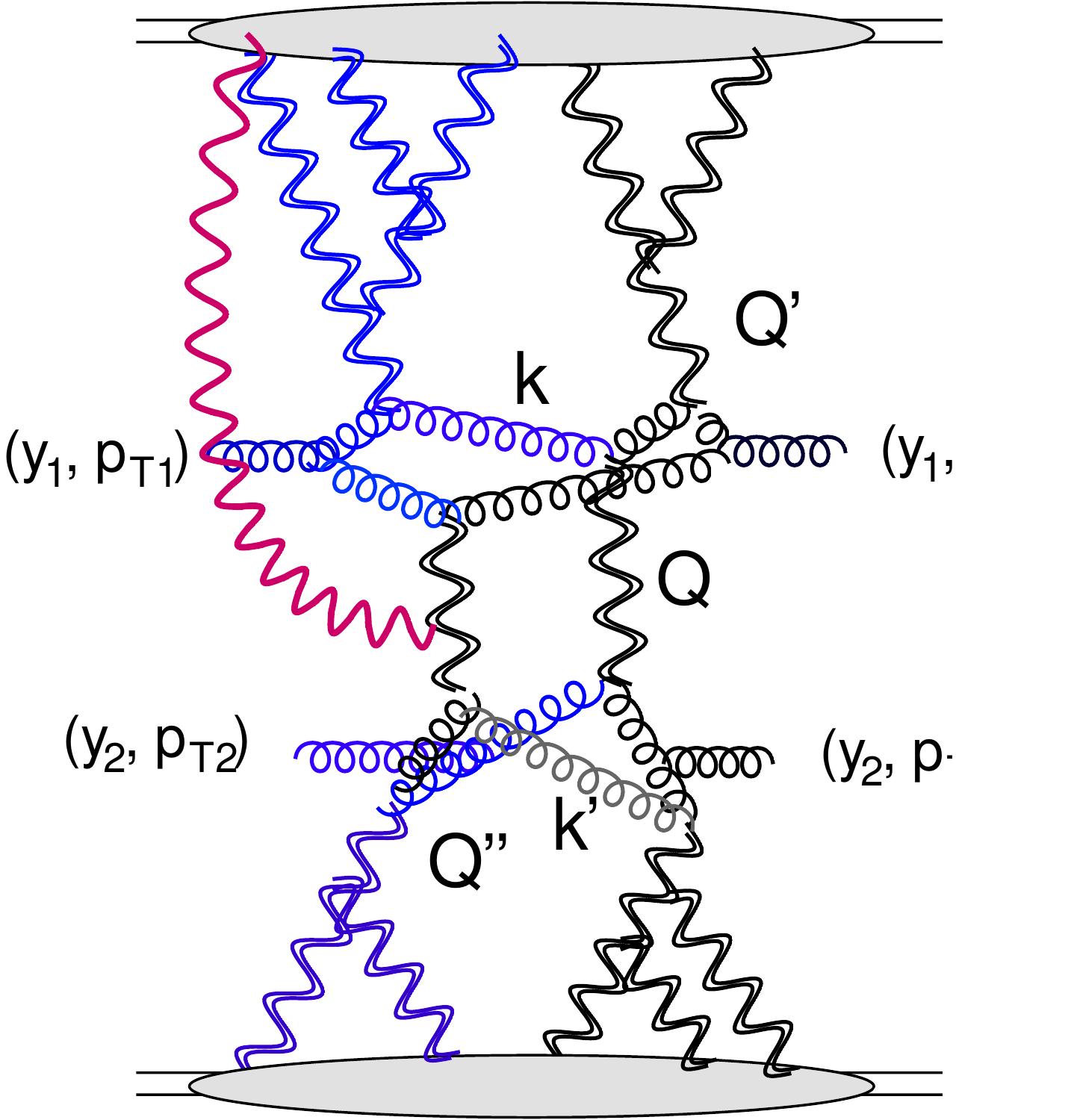}  \\
      \fig{DN}-a &  & \fig{DN}-b\\
      \end{tabular}
          \caption{\fig{DN}-a:$-\Delta N\Lb\tau\Rb  = -4\,\tau \frac{1}{\tau}
\frac{d}{d \tau} \tau \frac{d}{d \tau}N\Lb \tau\Rb$  versus 
      $\tau$ for the behavior of the scattering amplitude deep in the
 saturation domain\cite{LETU}.  \fig{DN}-b shows the example of a more
 complicated structure of the partonic cascades than the exchange of
 the BFKL Pomeron, which are shown in \fig{mudia}. The colour of the
 lines indicates  the parton shower.}     
\label{DN}
   \end{figure}

  At first sight   \eq{AM2} shows that  the amplitude   has a maximum 
 at $\tau = r^2_1 \,Q^2_s=1$. However, this is not correct. \eq{AM2} 
gives the correct behavior for $\tau < 1$ while for $\tau >1$ we need to
 take into account the interaction of the BFKL Pomerons and the non-linear
 evolution, generated  by these interactions. The general 
  result of this evolution is the fact that the amplitude depends on one
 variable\cite{GS} $\tau$, i.e.  $N\Lb \tau\Rb$ (as it has geometric 
scaling behavior).
 The peak at $\tau=1$ appears in 
  \beq \label{AM3}
  \nabla^2_{r_1} N\Lb r_1,r_2; Y\Rb \,=\,\,4\,Q^2_s(Y) \frac{1}{\tau}\frac{d}{d \tau} \tau \frac{d}{d \tau}N\Lb \tau\Rb
  \eeq
  From \eq{AM3} we can conclude that the width of the distribution
 in $r^2_1$ is of the order of $Q^2_s$,
  but depends  crucially  on the model
 for the Pomeron interaction. In \fig{DN}-a we 
  plot this value for the behavior of the scattering amplitude deep
 in the saturation domain (see Ref.\cite{LETU}).
  
 This approach is not correct for $\tau \to 1$ and $- \nabla^2 N = 1.58 $
 at $\tau=1$,  but it starts to be small at $\tau > 2$, which could be 
large enough to trust the formulae of Ref.\cite{LETU}. At least such a 
conclusion can be justified considering the fit of the DIS data in the
 saturation model of Refs.\cite{CLP,CLP1}, which based on the idea of
 Ref.\cite{IIM}, and has the  correct behavior of the scattering 
amplitude,
 both  deep in the saturation domain, and near $\tau = 1$. Hence, we 
expect
 that $\nabla^2 N$ decreases faster than we can see from \eq{AM2}. Bearing
 these conclusions in mind, we will calculated the contribution of 
\fig{mudia},   keeping all $N$ in \eq{MAEQC1} in the vicinities of
 the saturation scales, by replacing  $\int^\infty_0  d \tau (-\nabla^2 
N)=
-\int^1_0  d \tau (-\nabla^2 N)$.
 
 We will  show in the folowing, that we cannot  take all six 
Pomerons in the
 vicinities of the saturation scale.  We have to take two of them,
 either deep in the saturation domain, or in the perturbative QCD region.
 We choose to  take two Pomerons between rapidities $y_1$ and $y_2$,
 (see \fig{DN}-b) i.e. in the perturbative QCD region. Unfortunately,
 we cannot use the AGK cutting rules\cite{AGK}, which state that these
 Pomerons will be not affected by the Pomeron interaction, and 
the contributions
 of these interactions (see red Pomeron in \fig{DN}-b) are canceled.
 Indeed, it has been proven that for the double inclusive 
production\cite{KOJA}, that
  they do not work in perturbative
 QCD. On the other
 hand, these Pomerons carry  transverse momentum $Q_T$ ,  unlike the others one in the diagram, which is
 larger than the saturation scale $Q_s\Lb y_2\Rb$ and therefore,
 their contributions are suppressed in comparison with the other Pomerons in \fig{mudia}.  In addition our choice provides the natural matching with the region $\bas \,y_{12}\,
   <\,1$.

    The integration over $Q_T$ will produce the same result as
 \eq{DISD7}, as in the previous section. We re-write the integration 
over $\phi_i$ (see \eq{DISD8}) in the following way:
   \beq \label{INTR}
 \int \prod d \phi_i\,e^{-i\, \vec{p}_{T1} \cdot \vec{\tilde{r}}'_1}\delta^{(2)}\Lb \vec{r}_1 + \vec{r}'_1 - \vec{\tilde{r}}_1 - \vec{\tilde{r}}'_1\Rb \,\to\, \Lb 2 \pi\Rb^4  \int d \phi \,e^{i \vec{p}_{T1} \cdot \vec{r}}    \int k_T d k_T\,J_0\Lb k_T r\Rb 
 \,J_0\Lb k_T r_1\Rb  \,J_0\Lb k_T \tilde{r}_1\Rb \,J_0\Lb k_T \tilde{r}'_1\Rb 
 \eeq   
   We see that the integrals over $r'_1$ and $r'_1$ leads to
 $r_1 \sim 1/Q_s(Y - y_1)$ and $r'_1 \sim 1/Q_s(Y - y_1)$.
   The same  holds for the integrals over  $r'_2$ and
 $r'_2,$    leading to    $r_2 \sim 1/Q_s(y_2)$ and $r'_2 \sim 1/Q_s(y_2)$ 
.
   Assuming that $Q_s\Lb Y - y_1\Rb \,>\,Q_s\Lb y_2\Rb$ we conclude
 that $ r_i$ and $r'_i$ are much smaller than $r_2$ and $r'_2$.
 Replacing
   \beq \label{AM4} 
    \nabla^2_{r_1} N_{\rm pr}\Lb r_1; Y - y_1\Rb\,e^{i \vec{Q}_T \cdot \vec{\tilde{r}'}_1}\,\nabla^2_{\tilde{r}'_1} N_{\rm pr}\Lb\tilde{r}'_1; Y - y_1\Rb  \,\to\,\frac{2^8\,\bar{\gamma}^4}{r_1\,r'_1} \Lb r^2_1\,Q^2_s\Lb Y - y_1\Rb\Rb^{\bar{\gamma}}     \Lb r'^2_1\,Q^2_s\Lb Y - y_1\Rb\Rb^{\bar{\gamma}} 
    \eeq
    where $\bar{\gamma} = 1 - \gamma_{cr}$, we obtain from \eq{INTR} that integration over $r$ takes the form
    \beq \label{AM5}
  \frac{1}{Q_s} \frac{1}{1 + 2\bar{\gamma}}\,\int^1_0 d \tau \, J_0\Lb\frac{k_T}{Q_s}\sqrt{\tau}\Rb\,\tau^{\bar{\gamma}}\frac{d \tau}{2 \sqrt{\tau}}\,=\, \frac{1}{Q_s} \frac{1}{1 + 2\bar{\gamma}}\,  {}_1F_2\Lb\{\h + \bar{\gamma}\},\{1, \frac{3}{2} + \bar{\gamma}\}, - \frac{k^2_T }{4 \,Q^2_s}\Rb 
   \eeq

  Recall that we consider $Q_s = Q_s(Y - Y_1) $ in \eq{AM5}. For
 $p_{T1} \ll\,Q_s\Lb Y - y_1\Rb$ we can replace 
  $e^{-i\, \vec{p}_{T1} \cdot \vec{\tilde{r}}'_1}\,=\,1$ in \eq{INTR}.
 In this case the integral  has the form
  
  \beq \label{AM51}
\frac{1}{Q^2_s} \frac{1}{(1 + 2\bar{\gamma})^2}\,\int^1_0  d \tau'\,\Bigg( \frac{1}{Q_s} \frac{1}{1 + 2\bar{\gamma}}\,  {}_1F_2\Lb\{\h + \bar{\gamma}\},\{1, \frac{3}{2} + \bar{\gamma}\}, - \frac{1 }{4 \,\tau'}\Rb \Bigg)^2\frac{d \tau'}{\tau'}\,=\,0.18/Q^2_s
  \eeq
  where $\tau' = k^2/Q^2_s$.

  The integral over  $r$ in the  lower part of the diagram takes the 
form:
  \beq \label{AM6}
     \int \frac{d^2 r'}{r'^2} \,J_0\Lb k_T \, r\Rb \,=\, \pi \,\ln\Lb k^2_T/(4 \mu^2_{\rm soft})\Rb
     \eeq
  Using \eq{AM6} for $p_{T2} \ll Q_s\Lb y_2\Rb  $ the integral
 over $k'_T$ can be reduced to
  \beq  \label{AM61}   
 \frac{1}{(1 + 2\bar{\gamma})^2}\,\int^1_0  d \tau''\,\Bigg( \,
  {}_1F_2\Lb\{\h + \bar{\gamma}\},\{1, \frac{3}{2} + \bar{\gamma}\}, - \frac{1 }{4 \,\tau''}\Rb \Bigg)^2\,\Lb \frac{\ln\Lb \tau''\Rb}{\tau''}\Rb^2\,=\,3.50
 \eeq

   Finally, collecting all numerical coefficients, we obtain
      \bea \label{DISD15}   
  &&   \frac{d^2 \sigma}{d y_1 d^2 p_{T1}\,d y_2 d^2 p_{T2}}\Lb \fig{DN}-b\Rb\,\,= \\
  &&   \,\,\mbox{C}^4\,2^3 \pi^3 \Lb\frac{2 C_F }{\as\,(2 \pi)^2 }\Rb^2\,\,\frac{1}{p^2_{T1}\,p^2_{T2}} \frac{0.18\,3.5}{Q^2_s\Lb Y - y_1\Rb}
 \Lb 2 \bar{\gamma}\Rb^8 \, \sqrt{\frac{1}{\Lb 2 \,D\,y_{12}\Rb^3}}\,\,e^{  2 \Delta_{\rm BFKL} \,y_{12}}\, \nn             \eea     
   where  constant $\mbox{C}$ is the value of the amplitude at $\tau=1$.
   
    This contribution is proportional to
   $$
     \propto ~~~e^{  2 \Delta_{\rm BFKL} \,y_{12}}\Big{/} Q^2_s\Lb Y-  y_1\Rb
     $$
     for $p_{T1} \ll Q_s\Lb Y - y_1\Rb$ and $p_{T2} \ll Q_s\Lb  y_2\Rb  $.
 Note that $Q^2_s\Lb Y - y_1\Rb \,>\, Q^2_s\Lb y_2\Rb$.
     
     We need to estimate the diagram of \fig{becor}-a (see \eq{MAIN}).
 This diagram can be re-written as
     \bea \label{DISD16}
        \frac{d^2 \sigma}{d y_1 d^2 p_{T1}\,d y_2 d^2 p_{T2}}\Lb \fig{becor}-a\Rb \,\,&=&\,\,\tilde{\mu}^2_{\rm soft}
             \frac{d^2 \sigma}{d y_1 d^2 p_{T1}} \Lb Q_T = 0; \eq{SINCL3}\Rb\,     \frac{d^2 \sigma}{d y_2 d^2 p_{T2}} \Lb Q_T = 0; \eq{SINCL3}\Rb \nn\\
   &    \mbox{where} & ~~~~\tilde{
             \mu}^2_{\rm soft}\,=\,\int d^2 Q_T\, N^2\Lb Q_T\Rb;
             \eea
                 
  Examining \eq{SINCL3}, one can see that in general case
 when $Y - y_1 \neq y_1$ and $ Y - y_2 \neq y_2$ all four
 Pomerons cannot be in the vicinity of the saturation scale.
 Actually we have two kinematic regions which give the maximal
 contributions (assuming $Q_s\Lb Y - y_1\Rb\,>\,Q^2_s\Lb y_1\Rb$):
  \begin{enumerate}
  \item \quad\,\,\,\,\, $ r^2 \,Q^2_s\Lb Y - y_1\Rb \,\approx \,1 $ but $r^2 Q^2_s\Lb y_1\Rb \to Q^2_s\Lb Y - y_1\Rb\Big{/}
       Q^2_s\Lb y_1\Rb \,\ll\,1$;
  \item \quad \,\,\,\,\,$ r^2 \,Q^2_s\Lb  y_1\Rb \,\approx \,1 $ but $r^2 Q^2_s\Lb y_1\Rb \to Q^2_s\Lb  y_1\Rb\Big{/}
       Q^2_s\Lb Y - y_1\Rb \,\gg\,1$;
       \end{enumerate}
In the region 1 the upper Pomeron is in the vicinity of the saturation
 scale, while the lower  Pomeron is in the perturbative QCD region. 
 In region 2 the lower Pomeron is in the vicinity of the saturation scale,
 and the upper Pomeron is deep inside the saturation domain.        
  As we have discussed (see \fig{DN}-a) $\nabla^2 N$ decreases in the
 saturation region much faster than in the perturbation QCD region and,
 therefore, we {bf assume} that the kinematic region 1 gives the largest
 contribution.        Hence, for $p_{T1}\, \ll\,Q^2_s\Lb y_1\Rb$
we obtain
\bea \label{DISD17}
 &&\frac{d^2 \sigma}{d y_1 d^2 p_{T1}} \Lb Q_T = 0; \eq{SINCL3}\Rb = \nn\\
 && \,\,\frac{8 C_F }{\as\,(2 \pi)^2} \frac{1}{p^2_T}\int d^2 r \,e^{i \vec{p}_T \cdot \vec{r}}\,\nabla^2 _r\,N^{\rm \tiny BFKL}_{\rm pr}\,\Lb r, r_1; Y- y, Q_T=0\Rb \,\nabla^2 _r\,N^{\rm \tiny BFKL}_{\rm tr}\,\Lb r, r_2; y, Q_T=0\Rb  \nn\\
 &&=\,\,\frac{8 C_F }{\as\,(2 \pi)^2} \frac{1}{p^2_T}
 \mbox{C}^2\,(4\bar{\gamma}^2)^2 \,\exp\Lb - \frac{ \ln^2\Lb Q^2_s\Lb Y - y\Rb/Q^2\Lb y \Rb\Rb}{4 D \,y}\Rb
 \eea
 In \eq{DISD17} we used  backward evolution, from the saturation 
boundary
 where $N = \mbox{C}$.

\begin{figure}[ht]
    \centering
  \leavevmode
  \begin{tabular}{ccc}
      \includegraphics[width=5.2cm]{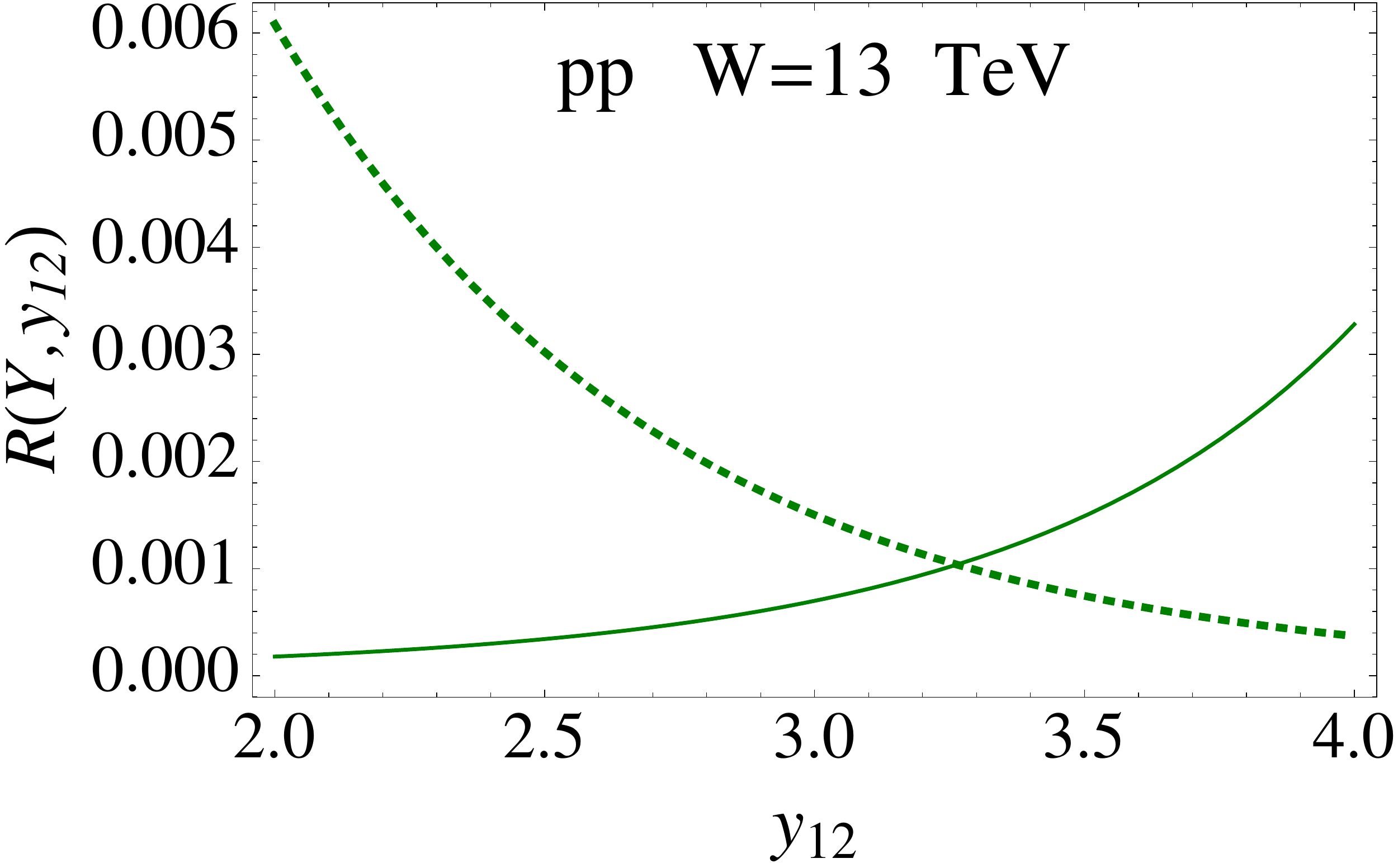}&  \includegraphics[width=5cm]{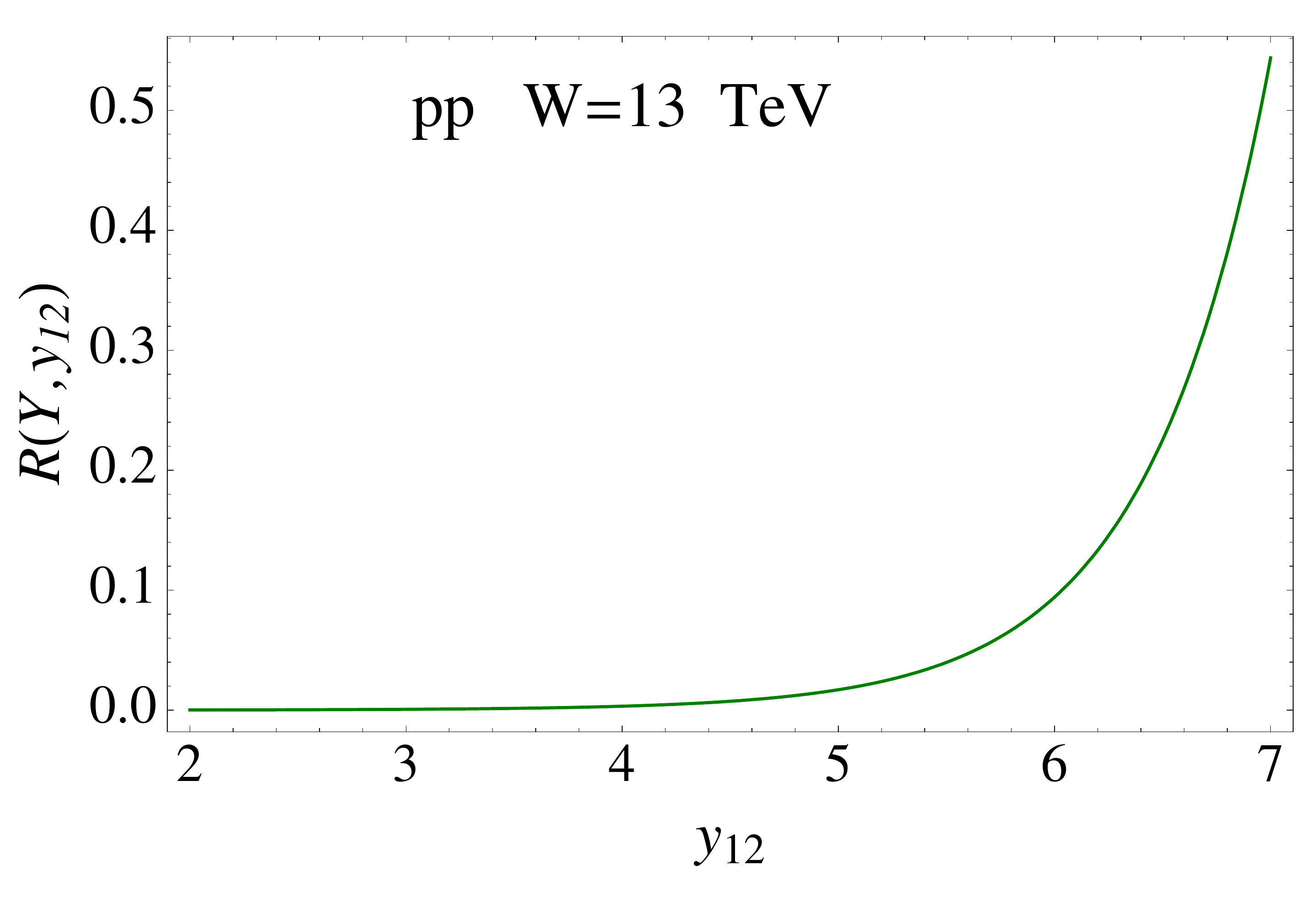}& \includegraphics[width=5.2cm]{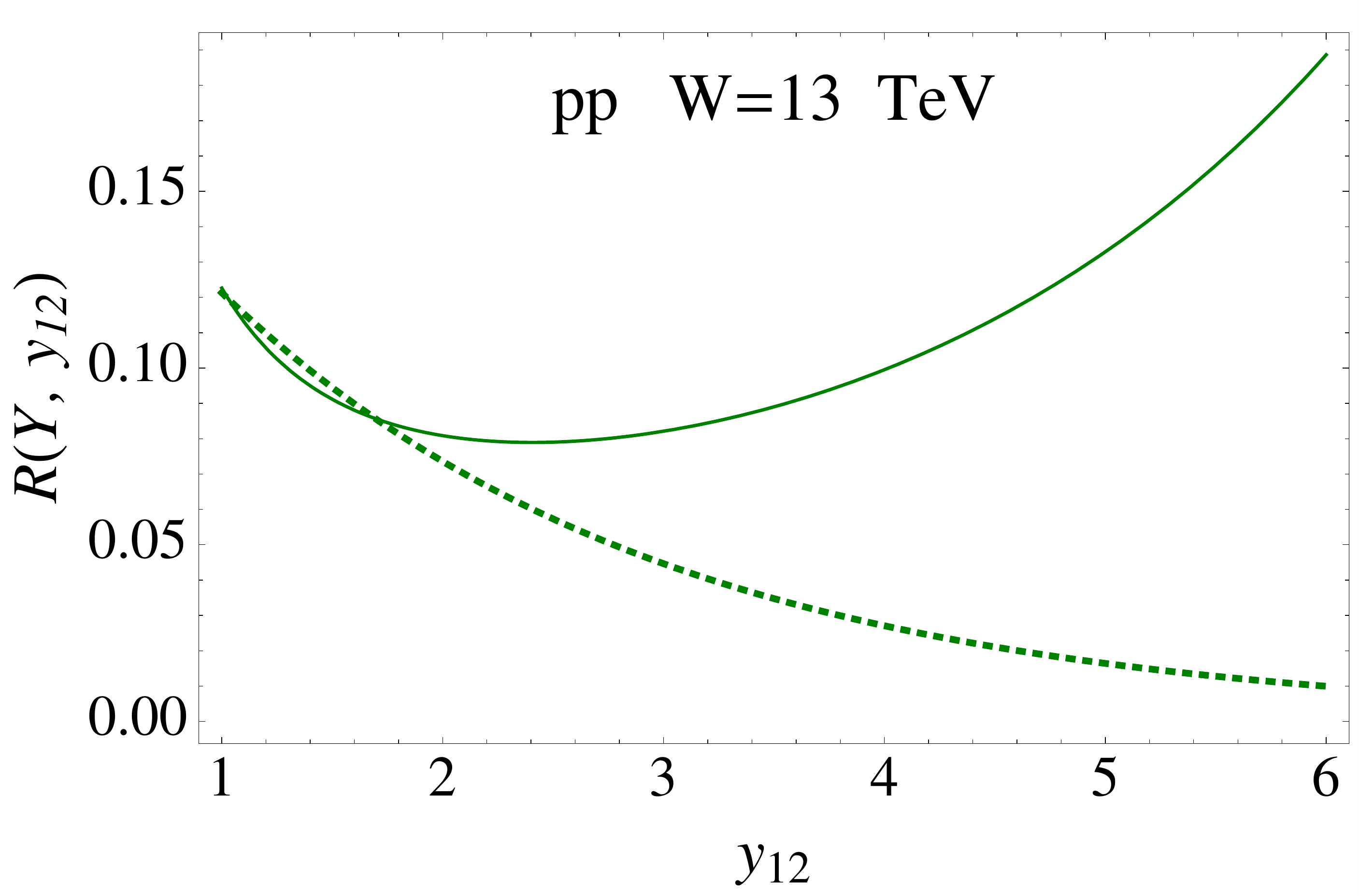}\\
      \fig{R}-a &    \fig{R}-b &\fig{R}-c\\
      \end{tabular}
      \caption{The ratio of \protect\eq{DISD18} at W=13 TeV versus
 $y_{12}$, assuming that the experiment has a symmetric pattern with
 $Y - y_1 = y_2 = \h(Y - y_{12})$. The dotted line in \fig{R}-a is
 the estimates for the    $y_{12}$ dependence of the Bose-Einstein
 contribution at small $y_{12}$ \protect\cite{GLMBEH,GOLELAST}. 
\fig{R}-a and \fig{R}-b give the estimates in the leading order of
 perturbative QCD with $\bas = 0.25$. In \fig{R}-c   we take
 $\Delta_{\rm BFKL} = 0.25  $ and $Q^2_s(Y) \propto \exp\Lb \lambda Y\Rb$
 with $\lambda = 0.25$. These numbers  correspond to the BFKL phenomenology.
      }
\label{R}
   \end{figure}

 The ratio of two contributions takes the following form:
 \bea \label{DISD18} 
&& R\,=\,\frac{\frac{d^2 \sigma}{d y_1 d^2 p_{T1}\,d y_2 d^2 p_{T2}}\Lb \fig{DN}-b\Rb }{\frac{d^2 \sigma}{d y_1 d^2 p_{T1}\,d y_2 d^2 p_{T2}}\Lb \fig{becor}-a\Rb} \,=\,\frac{1}{N^2_c - 1} \frac{\tilde{\mu}^2_{\rm soft}}{Q^2_s\Lb y_2\Rb}\,8\,\pi^5\,\Lb 2\,\bar{\gamma}\Rb^4 \,\,0.18\,\,3.5 \\
&&\times\,\, \,\sqrt{\frac{\pi}{\Lb 2 \,D\,y_{12}\Rb^3}}\,\,e^{  2 \Delta_{\rm BFKL} \,y_{12}}\,\,\exp\Lb  \frac{ \ln^2\Lb Q^2_s\Lb Y - y_1\Rb/Q^2\Lb y_1 \Rb\Rb}{4 D \,y_1}\Rb\exp\Lb \frac{ \ln^2\Lb Q^2_s\Lb Y - y_2\Rb/Q^2\Lb y_2 \Rb\Rb}{4 D \,y_2}\Rb\nn
\eea
 One can see that  \eq{DISD18} demonstrates the additional suppression in comparison with the calculation of the simplest diagram, due to infrared cutoff  at $Q_s\Lb y_2\Rb$ instead of $p_{T2}$. The factor $\exp\Lb  2 \Delta_{\rm BFKL} \,y_{12} \Rb$ reflects the fact that two BFKL Pomerons between rapidities $y_1$ and $y_2$ are  taken in the perturbative QCD region. It should be stressed that we can trust our estimates only  for values of $y_{12}$ at which the exchange of the BFKL Pomeron with rapidity $y_{12}$ give the contribution smaller than $ {\rm C}$. This condition means that 
  \beq \label{DISD19}
 \frac{1}{\Lb 2 \,D\,y_{12}\Rb}\,\,e^{  2 \Delta_{\rm BFKL} \,y_{12}} \,\,<\,\,{\rm C}
 \eeq 
Taking
 $\Delta_{\rm BFKL} = 0.25  $ and $Q^2_s(Y) \propto \exp\Lb \lambda Y\Rb$
 with $\lambda = 0.25$ (these values    correspond to the BFKL phenomenology) we obtain that the l.h.s. of \eq{DISD19}  is smaller than 0.15 for $y_{12}\, \leq \,7$. Therefore, we can trust our estimates  shown in \fig{R} for ${\rm C}\, > \,0.15$. We are used to take  ${\rm C}=0.3$ 
 which lead to the contribution of the shadowing corrections of the order of  30\%.
 
  Two last factors in \eq{DISD18}   stem from the perturbative QCD nature of two Pomerons in \eq{DISD16} ( see \eq{DISD17}).


 In \fig{R} we plotted ratio $R$ as function of $y_{12}$ {for $y_{12} \leq 7$ (see \eq{DISD19}}. One can see that
 the ratio increases for large $y_{12}$ .

  \section{Azimuthal angle correlations}
  
  
  The azimuthal angle correlations stem from terms $\Lb \vec{Q}_T
 \cdot \vec{r}_i\Rb^n$ in the vertices (see \eq{VSQ} and \eq{VSQ1}).
 Indeed, after integrating over $r_i$ these terms transform to
 expressions of the following type\cite{LERECOR}:\\ $ \Lb \vec{Q}_T
 \cdot \vec{p}_{T1}\Rb^{m_1}\,  \Lb \vec{Q}_T \cdot \vec{p}_{T2}\Rb^{m_2}$,
 which lead to term of $\Lb \vec{p}_{T1}  \cdot   \vec{p}_{T2}\Rb^m$.  We
 have illustrated in \eq{VSQ} and \eq{VSQ1} how these  originate
 from
 the general form of BFKL the Pomeron vertices in the coordinate 
representation.
  From \eq{VSQ} and \eq{VSQ1}   only  terms proportional 
to $
 \Lb \vec{Q}_T \cdot \vec{r}_{i}\Rb^n$ with even $n$ appear in the
 expansion. Therefore, the azimuthal angle ($\phi$)  correlation function
   contains only terms $\cos^{2 n}\Lb \phi \Rb$, and it is invariant with 
respect to $\phi \to \pi - \phi$. In  other words, $v_n$ with odd $n$, 
turn
 out to be zero. Hence, we have the first prediction: the value $v_n$ with
 odd $n$ should decreases with $y_{12}$, and their dependence should 
follow
 the dotted lines in \fig{R}-a.

 We return to \eq{DISD1} and integrate over $Q_T$, collecting
 terms that depend on the angles between $\vec{Q}_T$ and $\vec{r}_i$, 
which
 we have neglected in the previous section. As we have learned the typical
 values of $Q_T \propto 1/r_2 \sim 1/r'_2$ where $r_2$ and $r'_2$ are larger
 than $r_1$ and $r'_1$. In other words , we showed that  the main
 contributions stem from the kinematic regions: $ r^2_1 Q^2_s\Lb Y -
 y_1\Rb \sim 1$ ( $ r'^2_1 Q^2_s\Lb Y - y_1\Rb \sim 1$ ) and  
   $r^2_2 Q^2_s\Lb  y_2\Rb \sim 1$ ( $ r'^2_2 Q^2_s\Lb  y_2\Rb \sim 1$ ).
 Assuming that $Q_s\Lb Y - y_1\Rb \gg Q_s\Lb y_2\Rb$ we conclude that
 $r_1(r'_1)  \ll r_2(r'_2)$. The typical $Q_T$ is determined by the
 largest dipoles and, therefore, we expect  $Q_T \approx 1/r_2
 (1/r'_2)$, as has been  demonstrated above. Bearing these estimate
  in mind, we can replace vertices $V_{\nu_1}\Lb \vec{r}_1,
 \vec{Q}_T\Rb$ and  $V_{\nu_2}\Lb \vec{r}'_1, \vec{Q}_T\Rb$ 
 in \eq{DISD1}  by \eq{VSQ1}  in which we  put $Q_T = 1/r_2$ 
and $Q_T = 1/r'_2$, respectively.  Taking into account that
 $r_1/r_2\,\ll\,1 (r'_1/r'_2\,\ll\,1)$   we obtain
   \bea \label{ANGLECOR1}
V_{\nu_1}\Lb \vec{r}_1, \vec{Q}_T\Rb\,V_{\nu_2}\Lb \vec{r}'_1,
 \vec{Q}_T\Rb &=&\\
& &
\Bigg\{\left(\frac{r^2_1}{2^6}\right)^{-i \nu_1 } - \Lb \frac{Q^4_T r^2_1}{2^{6}}\Rb^{ i \nu_1}\Bigg\} \,\left( 1 \,-\,\frac{1}{2^4}  \Lb \vec{Q}_T \cdot \vec{r}_1\Rb^2     \,+\,\frac{1}{2^8} \Lb \vec{Q}_T \cdot \vec{r}_1\Rb^4 \right)\nn\\
  &\times&\, \Bigg\{\left(\frac{r'^2}{2^6}\right)^{-i \nu_2 } - \Lb \frac{Q^4_T r'^2_1}{2^{6}}\Rb^{ i \nu_2}\Bigg\} \,\left( 1 \,-\,\frac{1}{2^4} \Lb \vec{Q}_T \cdot \vec{r}'_1\Rb^2      \,+\,\frac{1}{2^8} \Lb \vec{Q}_T \cdot \vec{r}'_1\Rb^4 \right)  \nn 
      \eea   
 At first sight  \eq{ANGLECOR1} should enter two angles   
     between $\vec{Q}_T$ and $\vec{r}_1 $ and $\vec{r}'_1$, respectively.
 However, in the integrand  for integration over $r_i$ (see \eq{DISD8}) 
depends only on one vector $\vec{p}_{T1}$ . Therefore, after integration
 over all angles, we  find  that the angle $\phi$ in \eq{ANGLECOR1} is 
the
 angle between $\vec{Q}_T$ and $\vec{p}_{T1}$.
      
  For     vertices $V^*_{\nu_1}\Lb \vec{r}_2, \vec{Q}_T\Rb$ and 
 $V^*_{\nu_2}\Lb \vec{r}'_2, \vec{Q}_T\Rb$  in \eq{DISD1}  we 
  use \eq{VLQ}.    Finally, we need to  evaluate the integral
  \bea \label{ANGLECOR2}
  &&I_Q\,=\\
  &&\, -16 \,\nu_1\,\nu_2  \int Q_T d Q_T  \Bigg\{V_{\nu_1}\Lb \vec{r}_1, \vec{Q}_T\Rb\,V_{\nu_2}\Lb \vec{r}'_1, \vec{Q}_T\Rb\Bigg\}^{r_1\,=\,r'_1=1/Q_s\Lb Y - y_1\Rb}_{\eq{ANGLECOR1}}\!\!\!\!\!\!\!\!\!\!\!\!\!\!\!\!\!\!\!\!\!\!\!\! \Lb Q^2_T\Rb ^{ -i (\nu_1 +  \nu_2)} \frac{\cos^2\Lb \h \Qv \cdot \vec{r}_2\Rb}{Q^2_T \,r^2_2} \,\cos\Lb   \Qv \cdot \vec{r}_2\Rb\nn
    \eea
     with better accuracy that we did in section 5.1, keeping  the
dependence on the angle between $\vec{Q}_T$ and $\vec{r}_2$.  
 Note, that  the factor $   \cos\Lb   \Qv \cdot \vec{r}_2\Rb$ comes from $\exp\Lb
 i \, \Qv \cdot \vec{r}_2\Rb$ in \eq{MAEQC1}. Taking
 this integral we  substitute for the terms in parentheses in 
\eq{ANGLECOR1},
 $|Q_T| = 1/r^2_2 (1/r'^2_2)$.
    
    The integral is equal to 
    \bea \label{ANGLECOR3}
    I_Q &=&    2^{6  i (\nu_1 + \nu_2)} \Lb - 2^7\nu_1\,\nu_2\Rb\ \Lb \frac{r^2_1}{r^2_2}\Rb^{i\,(\nu_1 + \nu_2)}\frac{1}{r^2_2}    \\
    &\times& \,\left( 1 \,-\,\frac{1}{2^4} \frac{ \Lb \vec{n} \cdot \vec{r}_1\Rb^2 }{r^2_2}    \,+\,\frac{1}{2^8}\frac{ \Lb \vec{n} \cdot \vec{r}_1\Rb^4 }{r^4_2}\right) \,  \left( 1 \,-\,\frac{1}{2^4} \frac{ \Lb \vec{n} \cdot \vec{r}'_1\Rb^2 }{r'^2_2}    \,+\,\frac{1}{2^8}\frac{ \Lb \vec{n} \cdot \vec{r}'_1\Rb^4 }{r'^4_2}\right) \nn\\
    &\times& \Bigg\{ \frac{1}{ i (\nu_1 + \nu_2)} \,-\,\frac{9}{32} \,\cos\Lb 2 \phi_2\Rb\,+\,\frac{3}{16}
   \cos\Lb 4 \,\phi_2\Rb   \Bigg\}\nn
    \eea
    where $\vec{n} = \vec{Q}_T/Q_T$, and  $\phi_2$ is the angle between
 $\vec{n}$ and $\vec{n}_2 = \vec{r}_2/r_2$.  In
 \eq{ANGLECOR3} the terms in $\Lb \dots \Rb \Lb \dots \Rb$ stem
 from the expansion with respect to $r^2_1/r^2_2 \,\ll\,1$.
 However, for the terms in $\{\dots\}$
there are no such  small parameters, and we expand the
 function of $\phi_2$ in the Fourier series.
        
     Integrating over $\vec{n}$ one obtains
    \beq \label{ANGLECOR4}
      \Lb\dots\Rb     \Lb\dots\Rb      \left\{\dots\right\}\,=\,\frac{1}{ i (\nu_1 + \nu_2)} \,+\,\frac{3}{ 2^{10}} \frac{r^2_1}{r^2_2}\Lb\Lb\vec{n}_1 \cdot \vec{n}_2\Rb^2        \,+\Lb\vec{n}'_1 \cdot \vec{n}_2\Rb^2\Rb \,+\,\frac{3}{2^{12}} \frac{r^4_1}{r^4_2} \Lb\Lb\vec{n}_1 \cdot \vec{n}_2\Rb^4       \,+\,\Lb\vec{n}'_1 \cdot \vec{n}_2\Rb^4 \Rb   
      \eeq   
      where $\vec{n}_1 = \vec{r}_1/r_1,\vec{n}'_1 =
 \vec{r}_1/r_1$ and $\vec{n}_2 = \vec{r}_2/r_2$.     
 Deriving \eq{ANGLECOR4} we neglected the extra powers of
 $r^2_1/r^2_2$ , which are small.
 Finally
  \bea \label{ANGLECOR5}
    I_Q\Lb \vec{r}_1, \vec{r}',\vec{r}_2; \nu_1, \nu_2\Rb &=&    2^{6  i (\nu_1 + \nu_2)} \Lb - 2^7\nu_1\,\nu_2\Rb\ \Lb \frac{r^2_1}{r^2_2}\Rb^{i\,(\nu_1 + \nu_2)}\frac{1}{r^2_2}    \\
    &\times& \Bigg\{\frac{1}{ i (\nu_1 + \nu_2)} 
\,+\,\frac{9}{ 2^{10}} \frac{r^2_1}{r^2_2}\Lb\Lb\vec{n}_1 
\cdot \vec{n}_2\Rb^2        \,+\Lb\vec{n}'_1 \cdot \vec{n}
_2\Rb^2\Rb \,+\,\frac{3}{2^{12}} \frac{r^4_1}{r^4_2}
 \Lb\Lb\vec{n}_1 \cdot \vec{n}_2\Rb^4       \,+\,\Lb\vec{n}'_1 \cdot
 \vec{n}_2\Rb^4 \Rb  \Bigg\}  \nn \eea

   From \eq{MAEQC1} we can see that the integration over $r_i$ can be
  written in the form
   \bea \label{ANGLECOR6}
    &&\frac{d^2 \sigma}{d y_1 d^2 p_{T1}\,d y_2 d^2 p_{T2}}\Lb \fig{mudia}\Rb\,\,=\,\,\Lb\frac{2 C_F \,\mu^2_{\rm soft}}{\as\,(2 \pi)^2 }\Rb^2\,\,\frac{1}{p^2_{T1}\,p^2_{T2}}\,\\
 &&\times\,\,\int  d^2 r_1\,d^2  r'_1\, d^2 \tilde{r} _1\,d^2 \tilde{r}'_1  \,e^{-i\, \vec{p}_{T1} \cdot \vec{{r}}_1}\delta^{(2)}\Lb \vec{r}_1 + \vec{r}'_1 - \vec{\tilde{r}}_1 - \vec{\tilde{r}}'_1\Rb  \,\nabla^2_{\tilde{r}_1} \tilde{r}_1 V_{\rm pr}\Lb \tilde{r}_1\Rb\,\,\nabla^2_{\tilde{r}'_1}\tilde{ r}'_1 V_{\rm pr}\Lb\tilde{r}'_1\Rb \nn\\
 && \times \,\int  d^2 r_2\,d^2 r'_2\, d^2 \tilde{r} _2\,d^2 \tilde{r}'_2  \,e^{-i\, \vec{p}_{T2} \cdot \vec{{r}}_2}\delta^{(2)}\Lb \vec{r}_2 + \vec{r}'_2 - \vec{\tilde{r}}_2 - \vec{\tilde{r}}'_2\Rb  \,\nabla^2_{\tilde{r}_2} r_2 V_{\rm tr}\Lb\tilde{ r}_2\Rb\,\,\nabla^2_{\tilde{r}'_2} \tilde{r}'_2 V_{\rm tr}\Lb\tilde{r}'_2\Rb \nn\\
 &&\times\,\,
\nabla^2_{{r}_1} \nabla^2_{{r}_2}   \nabla^2_{{r}'_1} \nabla^2_{{r}'_2} \,\Bigg({r}_1\,{r}_2\,  {r}'_1\, {r}'_2\,  I_Q\Lb \vec{r}_1, \vec{r}',\vec{r}_2; \nu_1, \nu_2\Rb\Bigg)\nn\\
&&\times\,2 \pi \Bigg\{\frac{1}{ i (\nu_1 + \nu_2)} \,+\,\, \frac{9}{ 2^{10}}\frac{r^2_1}{r^2_2}\Lb\Lb\vec{n}_1 \cdot \vec{n}_2\Rb^2        \,+\Lb\vec{n}'_1 \cdot \vec{n}_2\Rb^2\Rb \,+\,\frac{3}{2^{12}} \frac{r^4_1}{r^4_2} \Lb\Lb\vec{n}_1 \cdot \vec{n}_2\Rb^4       \,+\,\Lb\vec{n}'_1 \cdot \vec{n}_2\Rb^4 \Rb  \Bigg\}  \nn
\eea 
             
  Each term in \eq{ANGLECOR6} can be factorized as a product of
 two functions  which depend on $r^i_1    $ and on $r^i_2$. 
Bearing this feature in mind we  calculate each
 term going to the momentum representation  using \eq{INTR}. 
   We obtain a product of  functions of $k_T$. Each of
 these function has the following general form:
  \beq    \label{ANGLECOR7}   
  \int d^2 r \,e^{ i \vec{k}_T \cdot \vec{r}}\,\prod_{i=1}^j\, r_{\mu_i}\,F\Lb r\Rb \,\,=\,\,(- i  \vec{ \nabla}_{k_T})^j \,\int d^2 r \,e^{ i \vec{k}_T \cdot \vec{r}}\,\,F\Lb r\Rb  \,\,=\,\,2 \pi\,(-i  \vec{ \nabla}_{k_T})^j \,\int d^2 r \,J_0\Lb k_T\,r\Rb\,\,F\Lb r\Rb  \eeq
   As we have seen the dependence on $\vec{r}_i$ stem from
 the integration over $Q_T$ or, in other words, from $I_Q$         
     In $I_Q$ dependence on $r_1   $ and $r'_1$  can be   
 extracted explicitly, leading to $F\Lb r\Rb \propto 1/r$.
 Hence the momentum image for \eq{ANGLECOR7} has a   simple form:
       \beq    \label{ANGLECOR8}   
  \int d^2 r \,e^{ i \vec{k}_T \cdot \vec{r}}\,\prod_{i=1}^j\, r_{\mu_i}\,F\Lb r\Rb \,\,=\,\,\,2 \pi\,(-i  \vec{ \nabla}_{k_T})^j \frac{1}{k_T}   
  \eeq       
  For $j =2$ and $j=4$ which we need to calculate \eq{ANGLECOR6}  we have
  \bea \label{ANGLECOR9}
&&(-i  \vec{ \nabla}_{k_T})^2 \frac{1}{k_T} 
  \,=\, \Big\{\frac{3}{k^5_T} \, k_{T,i} \, k_{T,i'} \,
-\,\frac{1}{k^3_T} \delta_{i,i'}\Big\};
\nn\\
&&(-i  \vec{ \nabla}_{k_T})^4 \frac{1}{k_T} \,=\, \Big\{\frac{105}{k^9_T} \, k_{T,i} \, k_{T,i'}  \, k_{T,j} \, k_{T,j'}\nn\\
&& \,-\,\frac{15}{k^7_T} \Lb \delta_{i j} \, k_{T,i'} \, k_{T,j'} \,+\,\delta_{i i'} \, k_{T,j} \, k_{T,j'} \,+\,\delta_{i' j} \, k_{T,i} \, k_{T,j'} \,+\,\delta_{j i'} \, k_{T,i} \, k_{T,j'}  \,+\,\delta_{i' j'} \, k_{T,j} \, k_{T,i} \,+\,\delta_{j j'} \, k_{T,i} \, k_{T,i'}\Rb\,\nn\\\,
&&+ \,\frac{3}{k^5} \Lb \delta_{i i'} \delta_{j j'}\,+\, \delta_{i j} \delta_{i' j'}\,+\,\delta_{i j} \delta_{i' j'}\Rb \Big\}; \nn
  \eea
Note that for integration over $\vec{r}_1$, \eq{ANGLECOR8} takes the form
\beq \label{ANGLECOR10}
 \int d^2 r_1 \,e^{ i (\vec{k}_T + \vec{p}_{T1}) \cdot \vec{r}_1}\,\prod_{i=1}^j\, r_{1,\mu_i}\,F\Lb r_1\Rb \,\,=\,\,\,2 \pi\,(-i  \vec{ \nabla}_{\vec{k}_T + \vec{p}_{T1}})^j \frac{1}{\sqrt{(\vec{k}_T\,+\,\vec{p}_{T1} )^2}}
\eeq
The term $\Lb r^2_1\Lb\vec{n}_1 \cdot \vec{n}_2\Rb^2      
  \,+ r'^2_1\Lb\vec{n}'_1 \cdot \vec{n}_2\Rb^2\Rb$
can be re-written as $ \Lb r_{1,\mu}\,r_{1,\nu}\,+\,r'_{1,\mu}
\,r'_{1,\nu}\Rb r_{2,\mu}\,r_{2,\nu}$ and in the momentum 
representation it looks as
\bea \label{ANGLECOR11}
&&\int d \phi \Bigg\{\Big(\frac{3}{k^5_T} \, k_{T,i} \, k_{T,i'} \,-\,\frac{\delta_{i i'}}{k^3_T}\Big) \frac{1}{\sqrt{k^2_T + p^2_{T1} + 2\cos\Lb \phi\Rb k_T p_{T1}}}\,+\\
&&\,\Big(\frac{3}{\Lb\sqrt{k^2_T + p^2_{T1} + 2\cos\Lb \phi\Rb k_T p_{T1}}\Rb^{5}} \, ( \vec{k}_{T} + \vec{p}_{T1})_i \,( \vec{k}_{T} + \vec{p}_{T1})_{i'}  \,-\,\frac{\delta_{i i'}}{\Lb\sqrt{k^2_T + p^2_{T1} + 2\cos\Lb \phi\Rb k_T p_{T1}}\Rb^3}\Big)\frac{1}{k_T}\Bigg\}\nn\\
&&=\,\,A \,\frac{p_{T1, i}\,p_{T1, i'}}{p^2_{T1}}\,+\, B\,\,\delta_{i i'} \nn
\eea

The expressions for $A$ and $B$ can be  written in a general form. 
 Assuming that both $p_{T1}$ and  $p_{T2}$ are smaller 
than $Q_s\Lb y_2\Rb$, we can expand the answer, only taking into
 account  terms that are proportional to $p^2_{T1}/k^2_T$
 and $p_{T2}^2/k'^2_T$. We obtain
\bea \label{ANGLECOR12}
A\Lb k_T, p_{T1}\Rb\,=\,\frac{3  p^2_{T1}}{4\,k^8_T}\Lb -13 k^2_T\,+\,50 \,p^2_{T1}\Rb ;~~~~~~~~~~~
B\Lb k_T, p_{T1}\Rb\,=\,\frac{1}{8\,k^4_T}\Lb 8 k^4_T + 65 k^2_T p^2_{T1} - 150 p^4_{T1}\Rb ; 
\eea

The integrations over $r'_2$ and $r_2$  differ from the integrations
 over $r_1$ and $r'_1$, due to extra factor $1/r^2_2$ which comes
 from the integration over $Q_T$ in \eq{DISD2} and \eq{DISD3}. 
Since $r^2_2 \approx \,1/Q^2_s(y_2)$ we replace it by $1/r^2_2 
= Q^2_s\Lb y_2\Rb$. In the case the integral over $k'_T$ takes the
 same form as  the integral over $k_T$, leading to the following
 expression which is  proportional to $\cos^2\Lb \phi\Rb$, where $\phi$ is
 the bangle between $\vec{p}_{T1}$ and $\vec{p}_{T2}$:
\beq \label{ANGLECOR13}
\frac{d^2 \sigma}{d y_1 d^2 p_{T1}\,d y_2 d^2 p_{T2}}\Lb \fig{mudia}\Rb\,
\,\propto\,\,\,Q^2_s\Lb y_2\Rb\,\,A\Lb k_T,p_{T1}\Rb\,A\Lb k'_T,p_{T2}\Rb\,
 \cos^2\Lb \phi\Rb
\eeq
which is responsible for the appearance of  $v_{2,2}$ and $v_2$.

Using  the second  expression in  \eq{ANGLECOR12} we can calculate the 
term which
 is proportional to $\cos^4\Lb \phi\Rb$  and  has the form
\beq \label{ANGLECOR14}
\frac{d^2 \sigma}{d y_1 d^2 p_{T1}\,d y_2 d^2 p_{T2}}\Lb \fig{mudia}\Rb\,\,\propto\,\,\,Q^2_s\Lb y_2\Rb\,\,A^{(4)}\Lb k_T,p_{T1}\Rb\,A^{(4)}\Lb k'_T,p_{T2}\Rb\, \cos^4\Lb \phi\Rb
\eeq
with
\beq \label{ANGLECOR15}
A^{(4)}\Lb k_T,p_{T1}\Rb\,\,=\,\,15\,\frac{573}{8}\frac{1}{k^6_T}\,\frac{p^2_{T1}}{k^2_T}\eeq

  The values of $v_2$ and   $v_4$ can be determined from
the following representation of the double inclusive cross section
  \beq \label{BEN1}
     \frac{d^2 \sigma}{d y_1 \,d y_2  d^2 p_{T1} d^2 p_{T2}}\,\,\propto\,\,1
 \,\,+\,\,2 \sum_n v_{ n,n } \Lb p_{T1}, p_{T2}\Rb \,\cos\Lb n\,
\varphi\Rb
     \eeq
     where $ \varphi$ is the angle between    $\vec{p}_{T1}$ and
 $ \vec{p}_{T2} $.
     $v_n$  is calculated from  $v_{n,n}   \Lb p_{T1}, p_{T2}\Rb $
     \beq \label{vn}  
 1.~~    v_n\Lb p_T\Rb\,\,=\,\,\sqrt{v_{n,n}\Lb p_T, p_T\Rb}\,;\,
~~~~~~~~~~~~~2.~~~~  v_n\Lb p_T\Rb\,\,=\,\,\frac{v_{n, n}\Lb p_T,
 p^{\rm Ref}_T\Rb}{\sqrt{v_{n,n}\Lb p^{\rm Ref}_T, p^{\rm Ref}_T\Rb}}\,;
     \eeq
 \eq{vn}-1 and \eq{vn}-2  depict   two methods  of how the  values
 of $v_n$ have been extracted from the experimentally measured
  $v_{n,n} \Lb p_{T1}, p_{T2}\Rb$, where $ p^{\rm Ref}_T$ denotes the
 momentum of the reference trigger.
     These two definitions are equivalent if  $v_{n, n}\Lb p_{T1},
 p_{T2}\Rb $ can be factorized as $v_{n, n}\Lb p_{T1}, p_{T2}\Rb\,=\,
     v_n\Lb p_{T1}\Rb\,v_n\Lb p_{T2}\Rb $. In this paper we use
 the definition in \eq{vn}-1.

 Introducing the angular correlation function as
  \beq \label{C} 
   C\Lb p_T, \phi\Rb \,\,\equiv\,\,\frac{\frac{d^2 \sigma}{d y_1 d^2 p_{T1}\,d y_2 d^2 p_{T2}}\Lb \fig{DN}-b\Rb }{\frac{d^2 \sigma}{d y_1 d^2 p_{T1}\,d y_2 d^2 p_{T2}}\Lb \fig{becor}-a\Rb}  
 \eeq
 we obtain
     
     \beq \label{BEN2}
     v_{n,n} \,\,=\,\,\frac{\int^{2 \pi}_0  d \phi \,C\Big(  \,p_T,\,\phi \Big)\,\cos\Lb n\,\phi\Rb}{2\,\pi\,+\, \int^{2 \pi}_0 
      d 
      \phi \,C\Big(  p_T, \phi\Big)}; ~~~~~~v_n\,=\,\sqrt{v_{n,n}}\,;
      \eeq

In \eq{DISD18} we have calculated  the part of $C\Lb p_{T}, \phi\Rb$
 which does not depend on $\phi$, which coincides with $C\Lb p_{T},
 \phi=0\Rb$ = $R$ of \eq{DISD18}  for $Q_s\Lb Y - y_1\Rb \,\gg\,
Q_s\Lb y_2\Rb$. To calculate the contribution to $C$, which depends
 on $\phi$, we need  to take the separate  integrals over $\nu_1$ and
 $\nu_2$ since the terms, which are proportional to $\cos^2\Lb \phi\Rb$
 and $\cos^4\Lb \phi\Rb$, do not have a pole at $\nu_1 = -\nu_2$ (see 
\eq{ANGLECOR5}). These integrations lead to the following extra factor
 in $C\Lb p_T,\phi\Rb\,-\,C\Lb p_T, \phi=0\Rb$
\bea \label{ANGLECOR14}
&&C\Lb p_T,\phi\Rb\,-\,C\Lb p_T, \phi=0\Rb\,\,\propto\,\,\,{\cal R}\,\frac{p^2_T}{Q^2_s\Lb Y-y_1\Rb}\,\frac{p^2_T}{Q^2_s\Lb Y-y_1\Rb}\,\,C\Lb p_T, \phi=0\Rb;\nn\\
&&{\cal R}\,=\,2\,\xi^2\,\,\sqrt{\frac{1}{\Lb 2\,D\,y_{12}\Rb^3}}\,\exp\Lb - 2 \xi^2/\Lb 4 \,D\,y_{12}\Rb\Rb
\eea
where $\xi \,=\,\ln\Lb Q^2_s\Lb Y - y_1\Rb/Q^2_s\Lb y_2\Rb\Rb$. 
We took factors proportional to $p_T$ from the expression for
 $ A\Lb k_T, p_{T1}\Rb$ and $A^{(4)}\Lb k_T, p_{T1}\Rb$ putting
 $p_{T1} = p_{T2} = p_T$.
To find the final correlation function and $v_{2,2}$ and
 $v_{4,4}$,
we need to collect  all numerical factors
 that come from $ A\Lb k_T, p_{T1}\Rb$ , $A^{(4)}\Lb k_T, p_{T1}\Rb$ 
and \eq{ANGLECOR6}, and to integrate over $\phi$, as  given  in
 \eq{C}.

 Note, that in the symmetric kinematics, where
 $ Y - y_1\,=\,y_2\,\,=\,\, \h \Lb Y - y_{12}\Rb$, $\xi = 0$ 
and \eq{ANGLECOR14} vanishes.  In this case, we have to use
 \eq{VSQ} instead of \eq{VSQ1}, keeping track of the corrections,
 which are proportional to $\nu_i$. As the result, we can consider
 $\xi = 0$ in \eq{ANGLECOR14}, but we need to replace factor $\xi^2 $ by 1.

 \eq{DISD18} and  \eq{ANGLECOR14} suffer la
 numerical uncertainties, which stem both from 
the values of soft parameters $\tilde{\mu}_{\rm soft}$ and $ \mu_{\rm soft}$
  as well as the values of the saturation scale at low energies, and from the
 integration in \eq{AM5} and  \eq{AM6}, which were taken neglecting
 contribution from the region  $\tau' \,<\,1$. On the other hand, the
 contribution to the double inclusive cross sections of the diagram of
 \fig{mudia} at $\bas \,y_{12}\,\ll\,1$ coincide with the contribution
 of \fig{becor}-b,
\beq \label{ANGLECOR15}
   \frac{d^2 \sigma\Lb \fig{mudia}\Rb}{d y_1 \,d y_2  d^2 p_{T1} d^2 p_{T2}}\,\,\,\xrightarrow{\bas \,y_{12} \to \,1}
   \,\,\,\,\,\frac{d^2 \sigma\Lb \fig{becor}-b\Rb}{d y_1 \,d y_2  d^2 p_{T1} d^2 p_{T2}}
   \eeq
   
   Therefore, to obtain the realistic estimate we use the following
 procedure of matching
   \bea \label{ANGLECOR15}
   v_2\Lb p_T = 5 \,GeV,  y_{12}  = 2\Rb|_{\fig{mudia}} \,\,&=&\,\,    v_2\Lb  p_T = 5 \,GeV  \Rb|_{\fig{becor}-b };\nn\\
 v_4\Lb p_T = 5 \,GeV, y_{12} = 2\Rb|_{\fig{mudia}} \,\,&=&\,\,    v_4\Lb  p_T = 5 \,GeV  \Rb|_{\fig{becor}-b };   
   \eea
   where $ v_2\Lb  p_T = 0.5 \,GeV  \Rb|_{\fig{becor}-b }$ and
 $ v_4\Lb  p_T = 0.5 \,GeV  \Rb|_{\fig{becor}-b }$ are taken 
from Ref.\cite{GLP} where the estimates were performed based
 on the model for soft interaction which describes all features
 of soft interaction at high energy and provides the interface
 with the hard processes.


    \begin{figure}[ht]
    \centering
  \leavevmode
  \begin{tabular}{ccc}
      \includegraphics[width=7.2cm]{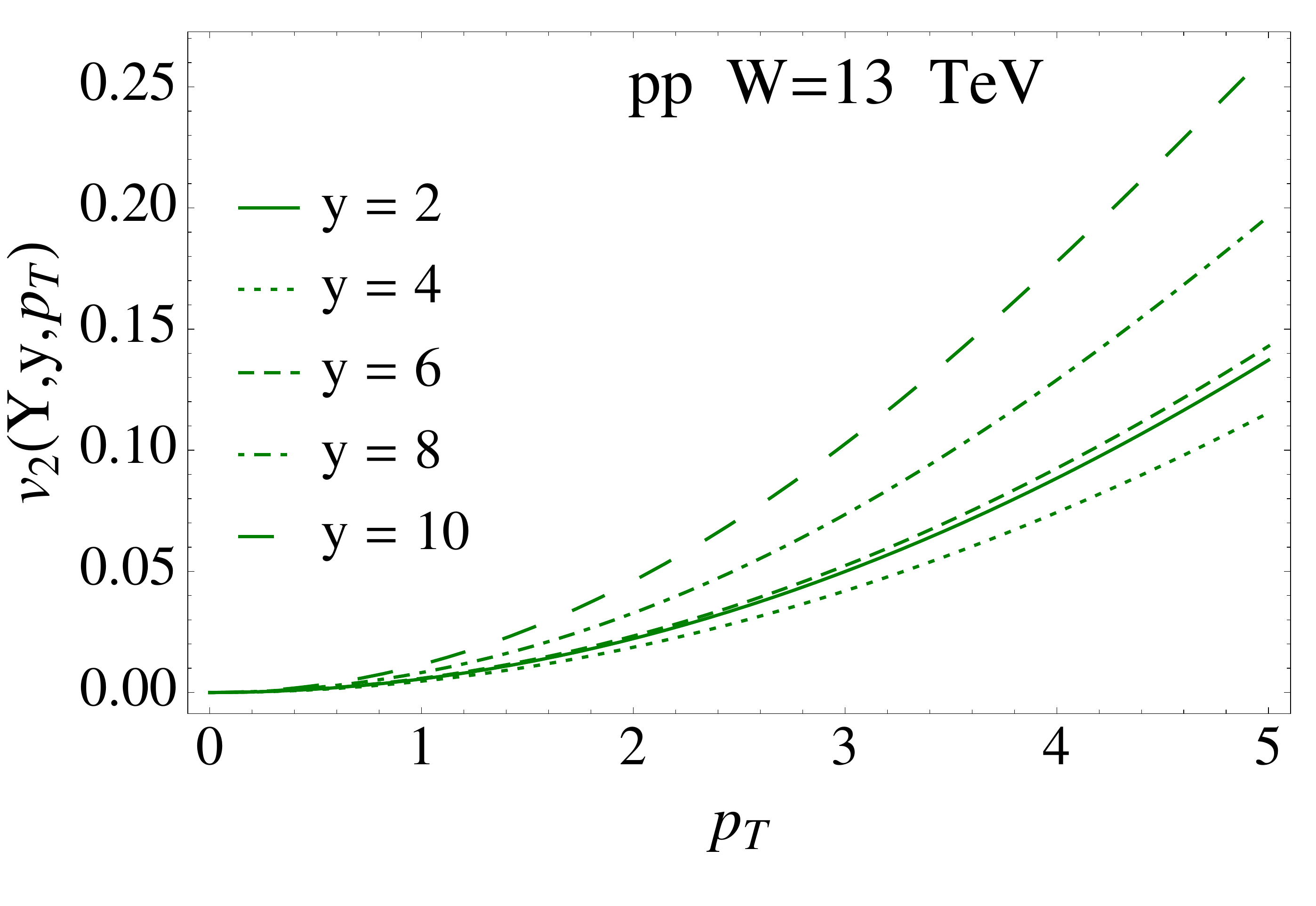}&  ~~~~~~~~~~~& \includegraphics[width=7.3cm]{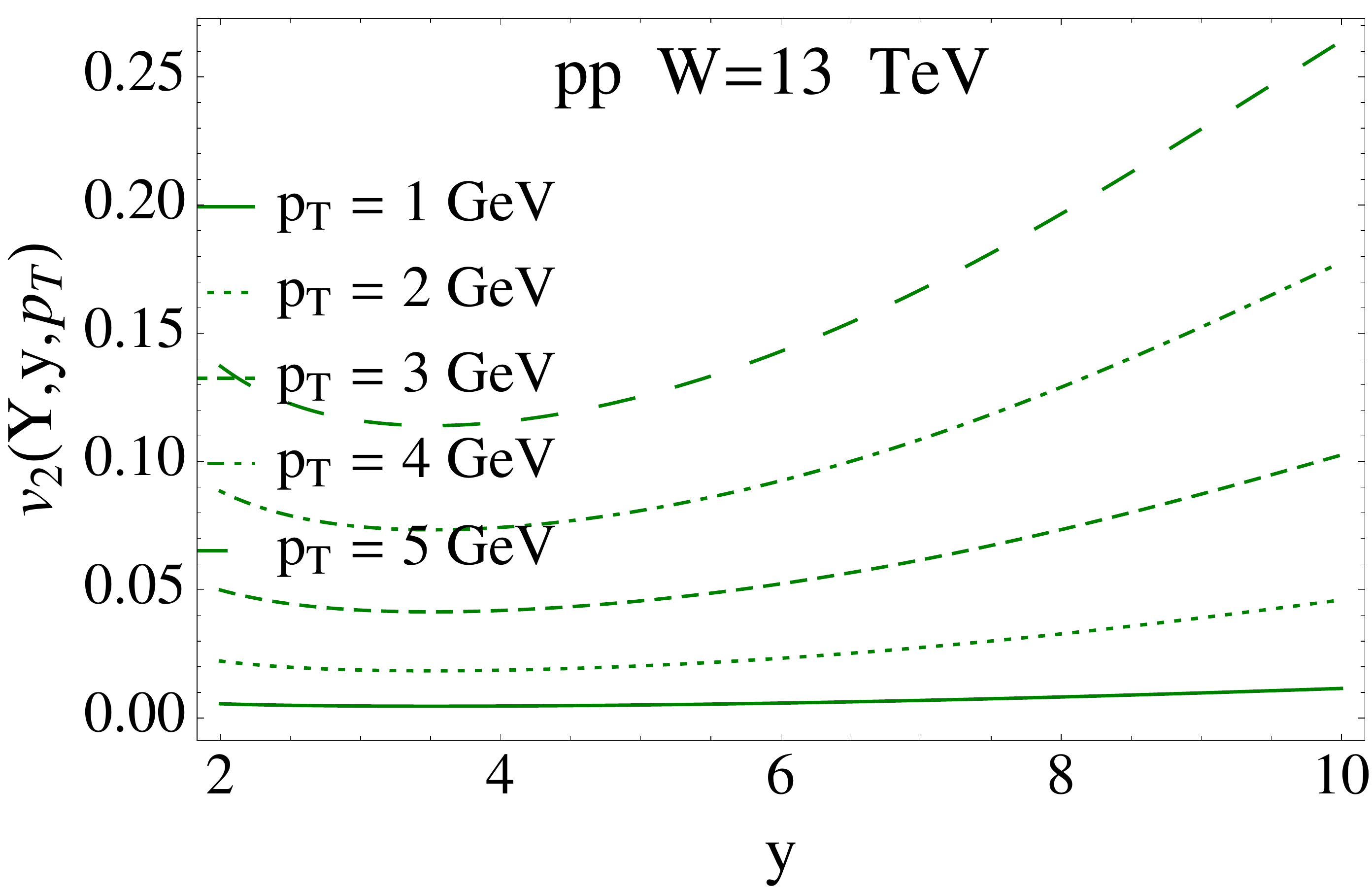}\\
      \fig{v}-a &   &   \fig{v}-b \\
       & &\\
          \includegraphics[width=7.2cm]{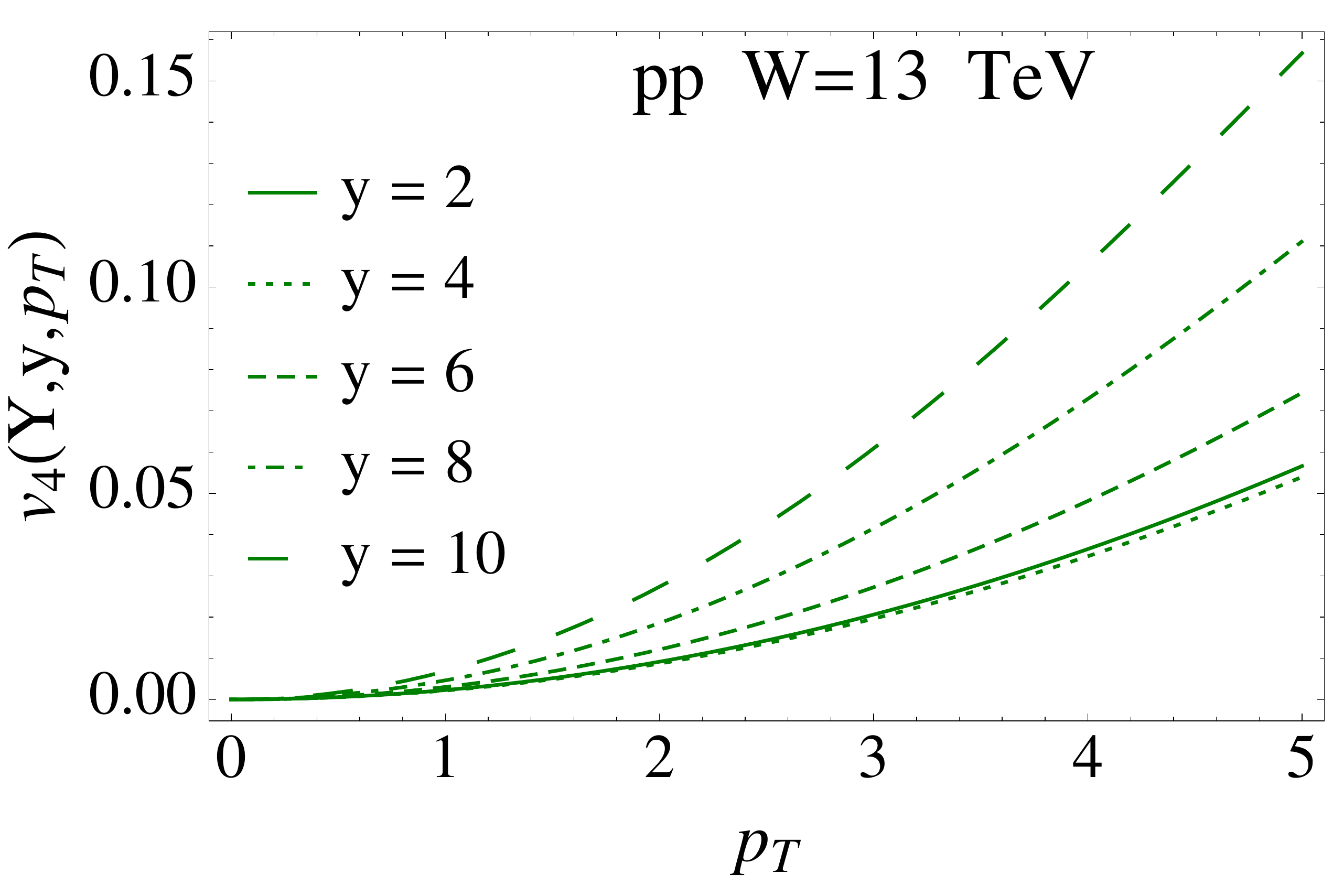}&  ~~~~~~~~~~~& \includegraphics[width=7.3cm]{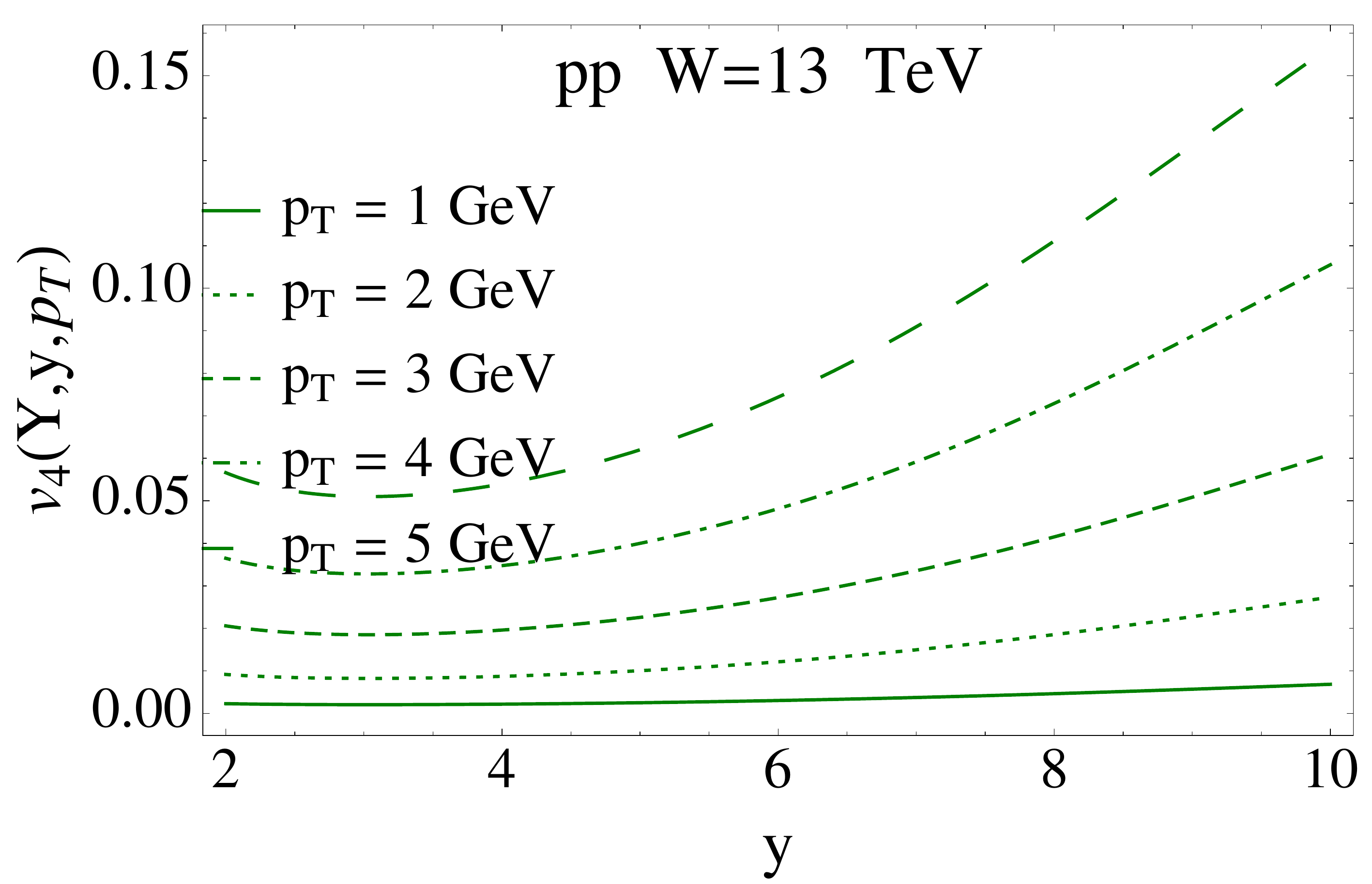}\\
      \fig{v}-c &   &   \fig{v}-d \\      
      \end{tabular}
      \caption{$v_n$ versus $p_T$ (\fig{v}-a and \fig{v}-c)  and versus $y$
 (\fig{v}-b and \fig{v}-d) at W=13 TeV  assuming that the experiment has a
 symmetric pattern with $Y - y_1 = y_2 = \h(Y - y_{12})$.  In all these
 figures we use \eq{ANGLECOR15}   for normalization  and    
      we take $\Delta_{\rm BFKL} = 0.25  $ and $Q^2_s(Y) \propto \exp\Lb 
\lambda Y\Rb$ with $\lambda = 0.25$. These numbers 
      correspond to the BFKL phenomenology. 
      }
\label{v}
   \end{figure}

   \fig{v} shows the $p_T$ and $y$ dependence of the $v_2$ and $v_4$
 using \eq{ANGLECOR15}   for normalization. In addition we take
 $\Delta_{\rm BFKL}$ = 0.25 and $Q^2_s( y)\, \propto\, \exp\Lb 
\lambda \,y\Rb$ with $\lambda = 0.25$. These values correspond
 to the BFKL Pomeron phenomenology. We believe that this figure
 illustrates the scale of rapidity dependence and will be
 instructive for future experimental observations.

  \section{Conclusions}
  
  In this paper we generalize the interference diagram, that described the Bose-Einstein correlation for small rapidity difference $ \bas y_{12}\,\ll\,1$, to include the emission of the gluons with rapidities ($y_i$)  between $y_1$ and $y_2$ ($ y_1\,,\,y_i\,<\,y_2$). We calculate the resulting diagram in CGC/saturation approach and make   
 two observations  which we consider as the main result of this paper.
 The first one is a substantial decrease of the odd Fourier
 harmonics $v_{2 n + 1}$ as a function of the rapidity difference
 $y_{12}$ ( see \fig{R}-c). The second result is, that even Fourier
 harmonic $v_{2 n}$ has a rather strong dependence on $y_{12}$,
 showing a considerable increase in the region of large $y_{12}$ 
(  see \fig{v}). We believe that our calculations, that have been
 performed both for the simplest diagrams and for the CGC/saturation
 approach, will be instructive for further development of the approach
 especially in the part that is related to the integration of the
 momenta transferred by the BFKL Pomerons. 
   
  We demonstrated in this paper the general origin of the density
 variation mechanism, whose nature does not depend on the technique
 that has been used. This mechanism has to be taken into account,
 since it leads to the values of Fouriers harmonics that are large
 and of the order of $v_n$ that have been observed experimentally. 
  
  We hope that the paper will be useful  in the clarification of the 
origin
 of the angular correlation, especially for hadron-hadron scattering at
 high energy.  We firmly believe that the experimental  observation of both phenomena: the sharp decrease of $v_n$ with odd $n$ and the substantial increase of $v_n$ with even $n$ as a function of $y_{12}$,  will be a strong argument for 
  CGC/saturation nature of the angular correlations.

  \section{Acknowledgements}
   We thank our colleagues at Tel Aviv university and UTFSM for
 encouraging discussions. Our special thanks go to  Carlos Contreras, 
 Alex Kovner and Misha Lublinsky for elucidating discussions on the
 subject of this paper.
 
  This research was supported by the BSF grant   2012124, by 
   Proyecto Basal FB 0821(Chile) ,  Fondecyt (Chile) grant  
 1140842, and by   CONICYT grant PIA ACT1406.  
 
 ~

\end{document}